\documentclass[12pt,a4paper]{article}
\usepackage[T1]{fontenc}
\usepackage[sc,osf]{mathpazo}
\usepackage{a4wide}  
\usepackage{amsfonts, amsthm}
\usepackage{amssymb}
\usepackage{amsmath}
\usepackage{bbm}
\usepackage{mathrsfs}
\usepackage{booktabs} 
\usepackage{ifpdf}
\DeclareMathOperator{\arccot}{arccot}
\ifpdf
\usepackage[pdftex,unicode,implicit]{hyperref}
\hypersetup{%
  pdftitle    = {On timelike supersymmetric solutions of gauged minimal
    5-dimensional supergravity}
  pdfkeywords = {supergravity, gauged supergravity, black holes, anti-de
    Sitter, unbroken supersymmetry, Kahler, isometry, holomorphic isometry, 
mono-holomorphic isometry, Gutowski-Reall
    black hole, Godel Universe, near-horizon geomtry},
  pdfauthor   = {Samuele Chimento, Tomas Ortin},
  plainpages  = true,
  colorlinks  = true,
  citecolor   = blue,
  urlcolor    = red,
  linkcolor   = black
}
\newcommand{\hepth}[1]{{\tt
\href{http://www.arXiv.org/abs/hep-th/#1}{hep-th/#1}}}

\newcommand{\arxiv}[1]{{\tt arXiv:\href{http://www.arXiv.org/abs/#1}{#1}}}
\newcommand{\doi}[1]{{\tt DOI:\href{http://dx.doi.org/#1}{#1}}}

\else
  \usepackage[dvips]{graphicx}
  \usepackage[unicode,implicit]{hyperref}
  \newcommand{\hepth}[1]{{\tt hep-th/#1}}

  \newcommand{\arxiv}[1]{{\tt arXiv:#1}}
  \newcommand{\doi}[1]{{\tt DOI:#1}}

\fi
\makeatletter
\@addtoreset{equation}{section}
\makeatother

\begin{document}

\begin{flushright}
\small
IFT-UAM/CSIC-16-094\\
\texttt{arXiv:yymm.nnnn [hep-th]}\\
November 28\textsuperscript{th}, 2016\\
\normalsize
\end{flushright}

\vspace{1.5cm}

\begin{center}

{\Large {\bf On timelike supersymmetric solutions of gauged minimal
    5-dimensional supergravity}}
 
\vspace{1.5cm}

\renewcommand{\thefootnote}{\alph{footnote}}
\renewcommand{\thefootnote}{\alph{footnote}}
{\sl\large  Samuele Chimento}\footnote{E-mail: {\tt Samuele.Chimento [at] csic.es}}
{\sl\large and Tom\'{a}s Ort\'{\i}n}\footnote{E-mail: {\tt Tomas.Ortin [at] csic.es}}

\setcounter{footnote}{0}
\renewcommand{\thefootnote}{\arabic{footnote}}

\vspace{1.5cm}

{\it Instituto de F\'{\i}sica Te\'orica UAM/CSIC\\
C/ Nicol\'as Cabrera, 13--15,  C.U.~Cantoblanco, E-28049 Madrid, Spain}\\ \vspace{0.3cm}

\vspace{1cm}


{\bf Abstract}

\end{center}

\begin{quotation}
  {\small We analyze the timelike supersymmetric solutions of minimal gauged
    5-dimen-\\sional supergravity for the case in which the K\"ahler base
    manifold admits a holomorphic isometry and depends on two real functions
    satisfying a simple second-order differential equation. Using this general
    form of the base space, the equations satisfied by the building blocks of
    the solutions become of, at most, fourth degree and can be solved by
    simple polynomic ansatzs. In this way we construct two 3-parameter
    families of solutions that contain almost all the timelike supersymmetric
    solutions of this theory with one angular momentum known so far and a few
    more: the (singular) supersymmetric Reissner-Nordstr\"om-AdS solutions,
    the three exact supersymmetric solutions describing the three near-horizon
    geometries found by Gutowski and Reall, three 1-parameter
    asymptotically-AdS$_{5}$ black-hole solutions with those three
    near-horizon geometries (Gutowski and Reall's black hole being one of
    them), three generalizations of the G\"odel universe and a few potentially
    homogenous solutions. A key r\^ole in finding these solutions is played by
    our ability to write AdS$_{5}$'s K\"ahler base space
    ($\overline{\mathbb{CP}}^{2}$ or SU$(1,2)/$U$(2)$) is three different, yet
    simple, forms associated to three different isometries.  Furthermore, our
    ansatz for the K\"ahler metric also allows us to study the dimensional
    compactification of the theory and its solutions in a systematic way.  }
\end{quotation}

\newpage
\pagestyle{plain}

\tableofcontents

\section*{Introduction}

The search for exact solutions of theories of gravity has been, and still is,
one of the most fruitful areas of work in gravitational physics. Symmetry has
probably been the main tool in this search and, therefore, it is not
surprising that, in gravity theories invariant under supersymmetry
transformations (theories of supergravity), unbroken supersymmetry has become
the main tool as well.\footnote{For a comprehensive review of supersymmetric
  solutions of supergravity theories with many references see,
  \textit{e.g.}~Ref.\cite{Ortin:2015hya}.} 

Unbroken supersymmetry is, indeed, a very powerful tool because, beyond the
fact that it implies the existence of ordinary symmetry (standard isometries
of the metric which also leave invariant the matter fields), relates in
non-trivial ways all the fields of the theory and, in particular, it relates
all the bosonic matter fields to the metric. This implies that all the fields
of a given solution with unbroken supersymmetry
(\textit{a.k.a.}~supersymmetric or BPS solution) can be constructed from a
common set of building blocks (functions, 1-forms, metrics in some submanifold
that satisfy simple equations or geometrical conditions) using different
combinations or rules. These combinations and rules are characteristic of each
supergravity theory and, identifying them, the building blocks and conditions
they satisfy makes it possible to construct large families of interesting
solutions and discover properties which cannot manifest themselves in single
members of the family.  The attractor mechanism
\cite{Ferrara:1995ih,Strominger:1996kf,Ferrara:1996dd,Ferrara:1996um,Ferrara:1997tw}
is, perhaps, the best known example of this kind of properties and their
relevance: only the knowledge of families of black-hole solutions with
different charges and values of the scalars at infinity can one realize that
their near-horizon values (and, hence, the entropy formulae) only depend on
the charges. The latter being quantized, a microscopic interpretation of the
entropy is, in principle, possible.

The systematic characterization or ``classification'' of supersymmetric
solutions was pioneered by Gibbons and Hull Ref.~\cite{Gibbons:1982fy} and,
specially, by Tod Ref.~\cite{Tod:1983pm} who showed that the requirement of
existence of just one unbroken supersymmetry in pure $\mathcal{N}=2,d=4$
supergravity was strong enough to identify a reduced number of building blocks
satisfying simple equations in terms of which all the components of the fields
of the supersymmetric solutions could be written. Shortly, Kowalski-Glikman
found all the solutions of the same theory admitting the maximal number of
unbroken supersymmetries (that is: 8) in
Ref.~\cite{KowalskiGlikman:1985wi}. 

However, since most of the solutions found by Tod were already
known\footnote{The bosonic sector of pure, ungauged, $\mathcal{N}=2,d=4$
  supergravity is the well-known and much studied Einstein-Maxwell
  theory. Then, it is no surprise that, for instance, the timelike
  supersymmetric solutions corresponded to the Perj\'es-Israel-Wilson family
  \cite{Perjes:1971gv,Israel:1972vx} which, as proven by Hartle and Hawking in
  Ref.~\cite{Hartle:1972ya}, only contains as regular non-trivial subfamily
  the Majumdar-Papapetrou solutions \cite{Majumdar:1947eu,kn:P} which describe
  extremal Reissner-Nordstr\"om black holes in static equilibrium.} and he
worked using the Newman-Penrose formalism, it was not until it was realized
that the Killing spinor equations could be rewritten as equations on tensors
constructed as spinor bilinears (a language much better understood by the
superstring community) that this line of research took off. This method was
successfully applied to the complete characterization of the supersymmetric
solutions of minimal 5-dimensional supergravity in
Ref.~\cite{Gauntlett:2002nw} leading to the discovery of a host of new and
interesting solutions. This procedure was immediately applied to ever more
complex cases. In the framework of $\mathcal{N}=2,d=4$ supergravity theories,
it was applied to

\begin{itemize}
\item Gauged, pure supergravity in Ref.~\cite{Caldarelli:2003pb}.
\item Ungauged but coupled to  vector multiplets in Ref.~\cite{Meessen:2006tu}.
\item Ungauged but coupled to vector multiplets and hypermultiplets in
  Ref.~\cite{Huebscher:2006mr}.
\item Coupled only to vector multiplets
  with Abelian gaugings in Refs.~\cite{Cacciatori:2008ek,Klemm:2009uw,Klemm:2010mc}.
\item Coupled to vector multiplets with non-Abelian gaugings (excluding
  SU$(2)$ Fayet-Iliopoulos terms)in Ref.~\cite{Hubscher:2008yz}.
\item Coupled to vector multiplets and hypermultiplets with the most general
  gauging (Abelian or not, with Fayet-Iliopoulos terms or not) in Ref.~\cite{Meessen:2012sr}
\footnote{Only the timelike supersymmetric solutions have
    been characterized in the most general case.}
\end{itemize}

In the $\mathcal{N}=1,d=5$ supergravity theories in which we are interested
here it has been applied to 

\begin{itemize}
\item Gauged, pure supergravity in Ref.~\cite{Gauntlett:2003fk}.
\item Coupled to vector multiplets with Abelian gaugings in
  Ref.~\cite{Gutowski:2004yv} for the timelike case (the results for the
  ungauged case were derived from those of the gauged one in
  Ref.~\cite{Gauntlett:2004qy}) and in Ref.~\cite{Gutowski:2005id} for the
  null case.
\item Ungauged but coupled to vector multiplets and hypermultiplets in
  Ref.~\cite{Bellorin:2006yr}.
\item Coupled to vector multiplets and hypermultiplets with the most general
  gauging in Ref.~\cite{Bellorin:2007yp}.
\item Coupled to vector and tensor multiplets and hypermultiplets with the
  most general gauging in Ref.~\cite{Bellorin:2008we}.
\end{itemize}

A feature of the 5-dimensional case, as compared with 4-dimensional one is
that, even in the simplest theory, some of the building blocks are not defined
by differential or algebraic equations but by geometrical conditions whose
general solution is not known. In particular, the most fundamental building
block of the 5-dimensional timelike supersymmetric solutions (which are the
ones we will be interested in here) is the so-called \textit{base-space
  metric}, which is a 4-dimensional Euclidean metric that enters in the
construction of the 5-dimensional spacetime metric and on which the
differential equations satisfied by the rest of the building blocks are
defined, is required to be hyperK\"ahler in the ungauged case (with no hypers,
as we will assume form now on to be the case) or just K\"ahler when there is
an Abelian gauging. These geometrical conditions are too general: we do not
know how to write a general 4-dimensional K\"ahler hyperK\"ahler metric in
terms of a set of functions, forms or lower-dimensional metrics satisfying
simple equations. This problem was solved in Ref.~\cite{Gauntlett:2002nw} by
considering only 4-dimensional hyperK\"ahler spaces admitting triholomorphic
isometries, which have Gibbons-Hawking metrics
\cite{Gibbons:1979zt,Gibbons:1979xm}, a constraint that still allows for many
interesting solutions like rotating and static
asymptotically-flat\footnote{The magic of the Gibbons-Hawking ansatz is that
  the additional isometry is compatible with spherically-symmetric
  (SO$(4)$-invariant) black-hole solutions and it does not restrict us to work
  with black strings.} (multi) black holes and black rings. These metrics are
defined by a single building block: a function harmonic in $\mathbb{E}^{3}$,
customarily called $H$, and, on them, the rest of the supersymmetry conditions
can be solved completely in terms of another three harmonic functions. As a
bonus, upon dimensional reduction along the additional isometry one finds
4-dimensional supersymmetric black holes.

The same ansatz has recently been used in theories of $\mathcal{N}=1,d=5$ with
vector multiplets and non-Abelian gaugings (but no Fayet-Iliopoulos terms), or
$\mathcal{N}=1,d=5$ Super-Einstein-Yang-Mills (SEYM) theories
\cite{Meessen:2015enl}. The general form of the timelike supersymmetric
solutions is a particular case of that found in Ref.~\cite{Bellorin:2007yp}
and the base space is also hyperK\"ahler. A piece of the non-Abelian 1-form
field is an anti-selfdual instanton on the hyperK\"ahler base space. If one
assumes that this space is Gibbons-Hawking one can then use Kronheimer's
results \cite{kn:KronheimerMScThesis} to solve the instanton equation on
that space in terms of BPS monopole
solutions to the Bogomol'nyi equation on $\mathbb{E}^{3}$
\cite{Bogomolny:1975de}. For the gauge group SU$(2)$ all the spherically
symmetric solutions of the Bogomol'nyi equation were found by Protogenov in
Ref.~\cite{Protogenov:1977tq} and one can profit from this result to construct
anti-selfdual instantons in the 4-dimensional hyperK\"ahler base
space.\footnote{Again, the magic of the Gibbons-Hawking ansatz is that the
  instantons built from the monopoles, which are only spherically symmetric in
  $\mathbb{E}^{3}$ (SO$(3)$) will be spherically symmetric in the
  4-dimensional base space if we make the simplest choice
  $\mathbb{R}^{4}_{\{0\}}$.} Somewhat surprisingly, the only monopoles that
give rise to regular instantons (the BPST one \cite{Belavin:1975fg}, in fact)
in the simplest setup belong to an intriguing class which has vanishing
asymptotic charge and a singularity at the origin and which give rise to
regular 4-dimensional non-Abelian black holes whose entropy, nevertheless,
depends on the non-Abelian field \cite{Bueno:2015wva}. These black holes were
called \textit{coloured black holes} in Ref.~\cite{Meessen:2008kb}. They exist
for more general gauge groups (because the corresponding \textit{coloured
  monopoles} also exist in more general gauge groups, as shown in
Ref.~\cite{Meessen:2015nla}) and are associated to 4-dimensional coloured
black holes in which the non-Abelian field configuration is the regular
instanton associated to the corresponding coloured monopole.

Given the success of this approach, it is a bit of a mystery that a similar
ansatz (\textit{i.e.}~assuming that the K\"ahler base space has a holomorphic
isometry) has not yet been used to simplify the Abelian-gauged case\footnote{It 
should be noted however that less general ansatzs have been used in the 
literature, namely toric \cite{Figueras:2006xx} and orthotoric 
\cite{Cassani:2015upa} K\"ahler base spaces. Actually many of 
the solutions we find here were already included in those works, either explicitly 
or as particular cases of more general solutions.}, which is
known to lead to complicated sixth-order differential equations
\cite{Gutowski:2004ez}. In that reference, Gutowski and Reall managed to find
a supersymmetric asymptotically-AdS$_{5}$ black-hole solution with a
squashed-$S^{3}$ near-horizon geometry plus two additional possible
non-compact near-horizon geometries. However, given the complexity of the
problem, they could not identify other supersymmetric asymptotically-AdS$_{5}$
black-hole solutions with the alternative near-horizon geometries.  Given the
connections between this kind of solutions and the AdS/CFT conjecture, finding
them constitutes an important open problem that could have been addressed by
making use of the aforementioned ansatz. Furthermore, as explained at the
beginning of this introduction, finding general families of solutions (or
extending the ones already known) is, by itself, an important goal. 

In a recent paper \cite{Chimento:2016run} we have shown how to write any K\"ahler metric
with a holomorphic isometry in a generalized Gibbons-Hawking form that depends
on just two real functions $H,W$ the first of which satisfies a $W$-deformed
Laplace equation on $\mathbb{E}^{3}$. In this paper we are going to use this
ansatz to simplify the equations and find more supersymmetric solutions of
minimal gauged 5-dimensional supergravity. We start by reviewing this theory
in Section~\ref{sec-thesugra} to introduce our notation and conventions. Its
bosonic sector is described in Section~\ref{sec-bosonicsector} and the
conditions found in Ref.~\cite{Gauntlett:2003fk} for a field configuration to
be a timelike supersymmetric solution will be reviewed in our notation in
Section~\ref{sec-timelikesusysolutions}. Then in
Section~\ref{sec-timelikesusysolutionswithisometry} we study the particular
case in which the base space of the timelike supersymmetric solution (a
4-dimensional K\"ahler space) has a holomorphic isometry, using the general
ansatz found in Ref.~\cite{Chimento:2016run}, finding a simpler set of equations to be
solved. Before we try to solve them, we have found it useful to rewrite in
Section~\ref{sec-examples} some well-known timelike supersymmetric solutions
(Reissner-Nordstr\"om-AdS$_{5}$ and AdS$_{5}$ itself) in a form and
coordinates adapted to our ansatz for the base space. We show three different
ways of writing AdS$_{5}$ in a timelike supersymmetric form, each of them
associated to a different form of writing the common base space
$\overline{\mathbb{CP}}^{2}$ or SU$(1,2)/$U$(2)$. In its turn, each of these
forms of AdS$_{5}$ will inspire a different ansatz for
asymptotically-AdS$_{5}$ timelike supersymmetric solutions. This will allow us
to obtain in Section~\ref{sec-solutions} two families of solutions
characterized by the parameter $\epsilon$ that constitute the main result of
this paper. The $\epsilon=1$ family, studied in Section~\ref{sec-e=1},
describes, among others, two kinds of solutions: asymptotically-AdS$_{5}$ rotating black
holes with the three possible near-horizon geometries found in
Ref.~\cite{Gutowski:2004ez} and the three near-horizon geometries as proper
timelike supersymmetric solutions.  The $\epsilon=0$ family, studied in
Section~\ref{sec-e=0}, describes a large number of
non-asymptotically-AdS$_{5}$ solutions of difficult interpretation. There are
three simple solutions in this class that are generalizations of the G\"odel
universe.  As in the ungauged and non-Abelian-gauged cases, all the solutions
found by using our ansatz can be immediately reduced to $d=4$ dimensions and
related to the solutions of some of the theories of $\mathcal{N}=2,d=4$
Abelian-gauged supergravity classified in
Refs.~\cite{Cacciatori:2008ek,Klemm:2009uw,Klemm:2010mc}. In the case we are
considering in this paper (minimal gauged supergravity), the corresponding
4-dimensional theory is the Abelian-gauged T$^{3}$ model and, in
Section~\ref{sec-reduction} we study the solutions of this theory that arise
from the 5-dimensional solutions discussed in the previous sections.
Section~\ref{sec-conclusions} contains our conclusions and directions for
future work. Finally, Appendices~\ref{app-3dmetric}, \ref{app-4dmetric} and
\ref{app-5dmetric} contain the connection and curvature of the 3-, 4- and
5-dimensional metrics that occur in this problem and Appendix~\ref{app-ads5}
contains a review of the construction of the AdS$_{5}$ metrics used in the
body of the paper.

\section{Minimal gauged $\mathcal{N}=1,d=5$ supergravity}
\label{sec-thesugra}

In this section we give a brief description of minimal gauged $\mathcal{N}=1,d=5$
supergravity and its timelike supersymmetric solutions. 

Minimal (\textit{pure}) $\mathcal{N}=1,d=5$ supergravity contains the
supergravity multiplet, only. This multiplet consists of the graviton
$e^{a}{}_{\mu}$, the gravitino $\psi^{i}{}_{\mu}$ and the graviphoton 1-form
$A_{\mu}$.  The spinor $\psi^{i}{}_{\mu}$ is a symplectic Majorana spinor and
$i$ is a fundamental $SU(2)$ (R-symmetry) index.\footnote{Our conventions are
  those in Refs.~\cite{Bellorin:2006yr,Bellorin:2007yp} which are those of
  Ref.~\cite{Bergshoeff:2004kh} with minor modifications.}

Since only one 1-form is available, and there are no scalars, at most a U$(1)$
subgroup of the SU$(2)$ R-symmetry group can be gauged. This is done by adding
a Fayet-Iliopoulos (FI) term $g n^{r}$, $r=1,2,3$ where $n^{r}$ is a constant
unitary vector which selects the $\mathfrak{u}(1)$ generator in
$\mathfrak{su}(2)$ that is going to be gauged: if $\{T_{r}\}$ are a basis of
the $\mathfrak{su}(2)$ Lie algebra, the generator of the U$(1)$ symmetry being
gauged will be $T\equiv n^{r}T_{r}$. $g$ is the gauge coupling constant and
only occurs in the bosonic action as a negative (AdS) cosmological constant as
we are going to see.

\subsection{The bosonic sector}
\label{sec-bosonicsector}

The bosonic action of minimal gauged 5-dimensional supergravity takes the form
of a cosmological Einstein-Maxwell theory supplemented by a Chern-Simons
term:

\begin{equation}
S  
=   
{\displaystyle\int} d^{5}x\sqrt{g}\
\biggl\{
R
+4g^{2}
-\tfrac{1}{4} F^{\mu\nu}F_{\mu\nu}
+\tfrac{1}{12\sqrt{3}}\frac{\varepsilon^{\mu\nu\rho\sigma\alpha}}{\sqrt{g}}
F_{\mu\nu}F_{\rho\sigma}A_{\alpha}
\biggr\}\, ,
\end{equation}

\noindent
where $F_{\mu\nu}=2\partial_{[\mu}A_{\nu]}$ and $g$ is the U$(1)$ coupling
constant. The cosmological constant $\Lambda$ is given in the above action
by\footnote{\label{foot:cosmocon} Our definition of the cosmological constant
  is such that it occurs in the $d$-dimensional Einstein-Hilbert action as
  \begin{equation}
  S=\int d^{d}x\sqrt{|g|} \left\{R -(d-2)\Lambda\right\}\, ,  
  \end{equation}
giving rise to the equations
\begin{equation}
G_{\mu\nu} = -\frac{(d-2)}{2}\Lambda g_{\mu\nu}\, ,
\,\,\,\,\,
\mbox{and}
\,\,\,\,\,
R_{\mu\nu} = \Lambda g_{\mu\nu}\, .  
\end{equation}
}

\begin{equation}
\Lambda = -\tfrac{4}{3}g^{2}\, ,  
\end{equation}

\noindent
and this value as well as the coefficient of the Chern-Simons term are fixed
by supersymmetry.

The equations of motion for the bosonic fields that follow from the above
action are

\begin{eqnarray}
G_{\mu\nu}
-\tfrac{1}{2}\left(F_{\mu}{}^{\rho} F_{\nu\rho}
-\tfrac{1}{4}g_{\mu\nu}F^{\rho\sigma}F_{\rho\sigma}
\right)      
-2g^{2}g_{\mu\nu}
& = &
0\, ,
\\ 
& & \nonumber \\
\nabla_{\nu}F^{\nu\mu}
+{\textstyle\frac{1}{4\sqrt{3}}} 
\frac{\varepsilon^{\mu\nu\rho\sigma\alpha}}{\sqrt{g}}
F_{\nu\rho}F_{\sigma\alpha}
& = &
0\, .
\end{eqnarray}

\subsection{Timelike supersymmetric configurations}
\label{sec-timelikesusysolutions}

The general form of the solutions of minimal, gauged, 5-dimensional
supergravity admitting a timelike Killing spinor\footnote{A timelike
  (commuting) spinor $\epsilon^{i}$ is, by definition, such that the real
  vector bilinear constructed from it $iV_{\mu}\sim
  \bar{\epsilon}_{i}\gamma_{\mu}\epsilon^{i}$ is timelike.} was found in
Ref.~\cite{Gauntlett:2003fk}. In what follows we are going to review it using
the notation and results of Ref.~\cite{Bellorin:2007yp} in which the most
general gauged theory was considered.

The building blocks of the timelike supersymmetric solutions are the scalar
function $\hat{f}$, the 4-dimensional spatial metric
$h_{\underline{m}\underline{n}}$,\footnote{$m,n,p=1,\cdots,4$ will be tangent
  space indices and $\underline{m},\underline{n},\underline{p}=1,\cdots,4$
  will be curved indices. We are going to denote with hats all objects that
  naturally live in this 4-dimensional space.} an anti-selfdual almost
hypercomplex structure $\hat{\Phi}^{(r)}{}_{mn}$,\footnote{That is: the 2-forms
  $\hat{\Phi}^{(r)}{}_{mn}$ $r,s,t=1,2,3$ satisfy
\begin{eqnarray}
\hat{\Phi}^{(r)\, mn} 
& = &
-\tfrac{1}{2}\varepsilon^{mnpq}\hat{\Phi}^{(r)}{}_{pq}\, , 
\hspace{1cm}
\mbox{or}
\hspace{1cm}
\hat{\Phi}^{(r)}=-\star_{4}\hat{\Phi}^{(r)}\, ,
\\
& & \nonumber \\
\hat{\Phi}^{(r)\, m}{}_{n}  \hat{\Phi}^{(s)\, n}{}_{p}
& = &
-\delta^{rs} \delta^{m}{}_{p} 
+\varepsilon^{rst}  \hat{\Phi}^{(t)\, m}{}_{p}\, .
\end{eqnarray}
} a 1-form $\hat{\omega}_{\underline{m}}$, and the 1-form potential
$\hat{A}_{\underline{m}}$. All these fields are defined on the 4-dimensional
spatial manifold usually called ``base space''. They are time-independent and
must satisfy a number of conditions:

\begin{enumerate}
\item The anti-selfdual almost hypercomplex structure $\hat{\Phi}^{(r)}{}_{mn}$, the
  1-form potentials $\hat{A}^{I}{}_{\underline{m}}$ and the base space metric
  $h_{\underline{m}\underline{n}}$ (through its Levi-Civita connection)
  satisfy the equation

\begin{equation}
\label{eq:dfAf}
\hat{\nabla}_{m}\hat{\Phi}^{(r)}{}_{np} 
+
g\varepsilon^{rst}n^{s}\hat{A}_{m}\hat{\Phi}^{(t)}{}_{np}
=0\, .
\end{equation}

\item The selfdual part of the spatial vector field strength
  $\hat{F}\equiv d\hat{A}$ must be related to the  function $\hat{f}$,
  the 1-form $\hat{\omega}$ by 

\begin{equation}
\label{eq:susy_{2}}
\hat{F}^{+} 
= 
{\textstyle\frac{2}{\sqrt{3}}} (\hat{f}d\hat{\omega})^{+} \, , 
\end{equation}

\item while the anti-selfdual part is related to the almost hypercomplex
  structure by

\begin{equation}
\label{eq:Fminus}
\hat{F}^{-} 
=
-2\hat{f}^{-1}n^{r}\hat{\Phi}^{(r)}\, .
\end{equation}

\item Finally, all the building blocks are related by the equation

\begin{equation}
\label{eq:susy_4}
\hat{\nabla}^{2}\hat{f}^{-1}
-\tfrac{1}{6}\hat{F}\cdot\hat{\star}\hat{F}
-{\textstyle\frac{1}{2\sqrt{3}}}
\hat{F}\cdot (\hat{f}d\hat{\omega})^{-}
=
0\, ,     
\end{equation}

\noindent
where the dots indicate standard contraction of all the indices of the tensors.

\end{enumerate}

Once the building blocks that satisfy the above conditions have been found,
the physical 5-dimensional fields can be built out of them as follows:

\begin{enumerate}
\item The 5-dimensional (conformastationary) metric is given by

\begin{equation}
\label{eq:conformastationaryd5}
  ds^{2} 
  = 
  \hat{f}^{\, 2}(dt+\hat{\omega})^{2}
  -\hat{f}^{\, -1}h_{\underline{m}\underline{n}}dx^{m} dx^{n}\, .
\end{equation}

\item The complete 5-dimensional 1-form field is given by

\begin{equation}
  \label{eq:completevectorfields}
  A 
  = 
  -\sqrt{3}\, \hat{f} (dt +\hat{\omega}) +\hat{A}\, ,    
\end{equation}

\noindent
so that the spatial components are

\begin{equation}
  A_{\underline{m}} 
  = 
  \hat{A}_{\underline{m}} -\sqrt{3}\hat{f} \hat{\omega}_{\underline{m}}\, ,
\end{equation}

\noindent
and the 5-dimensional field strength is

\begin{equation}
F
= 
-\sqrt{3} d[\hat{f} (dt +\hat{\omega})]  +\hat{F}\, .
\end{equation}

\end{enumerate}

As it has already been observed in Ref.~\cite{Gauntlett:2003fk}, from
Eq.~(\ref{eq:dfAf}) if follows that there is one complex structure
(generically given by $n^{r}\hat{\Phi}^{(r)}$) which is covariantly constant
in the base space

\begin{equation}
\label{eq:Kahlerstructure}
\hat{\nabla}_{m}(n^{r}\hat{\Phi}^{(r)}{}_{np}) =0\, ,
\end{equation}

\noindent
which, in its turn, implies that the base space metric
$h_{\underline{m}\underline{n}}$ is K\"ahler with respect to the complex
structure $\hat{J}_{mn}\equiv n^{r}\hat{\Phi}^{(r)}{}_{np}$ (see,
\textit{e.g.}~Ref.~\cite{kn:Joyce}).

It is convenient to choose, for instance, $n^{r}=\delta^{r}{}_{1}$. With this
choice, Eq.~(\ref{eq:dfAf}) splits into 

\begin{eqnarray}
\label{eq:df1=0}
\hat{\nabla}_{m}\hat{\Phi}^{(1)}{}_{np} 
& = &
0\, , 
\\
& & \nonumber \\
\label{eq:df2=pf3}
\hat{\nabla}_{m}\hat{\Phi}^{(2)}{}_{np} 
& = & 
g\hat{A}_{m}\hat{\Phi}^{(3)}{}_{np}\, ,
\\
& & \nonumber \\
\label{eq:df3=-pf2}
\hat{\nabla}_{m}\hat{\Phi}^{(3)}{}_{np} 
& = & 
-g\hat{A}_{m}\hat{\Phi}^{(2)}{}_{np}\, .
\end{eqnarray}

The first equation is just Eq.~(\ref{eq:Kahlerstructure}) for our particular
choice of FI term, which implies the choice of complex structure
$\hat{J}_{mn}\equiv\hat{\Phi}^{(1)}{}_{np}$.  Taking this fact into
account,\footnote{We use the integrability condition of Eq.~(\ref{eq:df1=0})
\begin{equation}
\hat{R}_{mnpq}=\hat{R}_{mnrs}\hat{J}^{r}{}_{p}\hat{J}^{s}{}_{q}\, ,    
\end{equation}
which leads to the relation between the Ricci and Riemann tensors
\begin{equation}
\hat{R}_{mn}=-\tfrac{1}{2}\hat{R}_{mprq}\hat{J}^{rq}\hat{J}^{p}{}_{n}\, .  
\end{equation}
The Ricci 2-form, defined as
\begin{equation}
\hat{\mathfrak{R}}_{mn}\equiv \hat{R}_{mp}\hat{J}^{p}{}_{n}\, ,  
\end{equation}
is, therefore, related to the Riemann tensor by 
\begin{equation}
\hat{\mathfrak{R}}_{mn} = \tfrac{1}{2}\hat{R}_{mnpq}\hat{J}^{pq}\, .  
\end{equation}
} the integrability condition of the other two equations is\footnote{If
  $gA_{m}$ vanishes (for instance, in the ungauged case), then we have a
  covariantly constant hyper-K\"ahler structure and, then, the base space is
  hyperK\"ahler.}

\begin{equation}
\label{eq:R=dP}
\hat{\mathfrak{R}}_{mn} 
=
-g\hat{F}_{mn}\, .  
\end{equation}

This equation must be read as a constraint on the 1-form potential
$\hat{A}_{\underline{m}}$ posed by the choice of base space metric.

Eq.~(\ref{eq:Fminus}) takes a simpler form as well:

\begin{equation}
\label{eq:susy_{3}}
\hat{F}^{-}
=
-2g\hat{f}^{-1}\hat{J}\, ,
\end{equation}

Tracing the first of these equations and Eq.~(\ref{eq:R=dP}) with
$\hat{J}^{mn}$ one finds a simple relation between the Ricci scalar of the
base space metric and the function $\hat{f}$:

\begin{equation}
\label{eq:RicciVf}
\hat{R}= 8g^{2}\hat{f}^{-1}\, .
\end{equation}

The last equation to be simplified by our choice is
Eq.~(\ref{eq:susy_4}). Substituting it in Eq.~(\ref{eq:susy_{3}}) one finds

\begin{equation}
\label{eq:susy_4-2}
\hat{\nabla}^{2}\hat{f}^{-1}
-\tfrac{1}{6}\hat{F}\cdot\hat{\star}\hat{F}
+\tfrac{1}{\sqrt{3}}g
\hat{J}\cdot (d\hat{\omega})
=
0\, . 
\end{equation}

\subsection{Timelike supersymmetric solutions 
with one additional isometry}
\label{sec-timelikesusysolutionswithisometry}

In order to make progress we need to make assumptions about the base space
K\"ahler metric so we can write it explicitly in terms of a small number of
functions that satisfy certain equations. In the ungauged
\cite{Gauntlett:2002nw,Bellorin:2006yr} and the non-Abelian gauged cases
\cite{Meessen:2015enl} in which the base space is hyper-K\"ahler it has proven
very useful to assume that the base space metric has an additional
triholomorphic isometry because, then, the metric is a Gibbons-Hawking metric
\cite{Gibbons:1979zt,Gibbons:1979xm} that depends on only one independent
function customarily denoted by $H$ which is harmonic in
$\mathbb{E}^{3}$. Writing the metric in terms of $H$ and other derived
functions simplifies the equations that depend on the metric so much that in
the ungauged case the complete solution can be written in terms of several
functions harmonic on $\mathbb{E}^{3}$.

It is natural to try the same strategy in the case at hands. We have shown in
Ref.~\cite{Chimento:2016run} that the most general K\"ahler metric admitting a
holomorphic isometry can be written as\footnote{\label{foot:JPhi}The
  associated complex structure has been chosen to be the anti-selfdual
\begin{equation}
\label{eq:J_matrix}
(J_{mn}) 
\equiv
\begin{pmatrix}
    \phantom{-}0_{2\times 2} & \phantom{-}\mathbb{1}_{2\times 2}\\
    -\mathbb{1}_{2\times 2}  & \phantom{-} 0_{2\times 2}
   \end{pmatrix}\, .
\end{equation}
We will identify it with $\hat{\Phi}^{(1)}$.
}

\begin{equation}
\label{eq:final_metric} 
ds^{2} 
= 
H^{-1}\left( dz+\chi \right)^{2}
+H\left\{(dx^{2})^{2}+W^{2}(\vec{x})[(dx^{1})^{2}+(dx^{3})^{2}]\right\}\, ,
\end{equation} 

\noindent
with the functions $H$ and $W$, and the 1-form $\chi$, depending only on the
three coordinates $x^{i}$ and satisfying the constraints

\begin{equation}
\label{eq:constraintijcurved}
\begin{array}{rcl}
(d\chi)_{\underline{1}\underline{2}} 
& = &
\partial_{\underline{3}}H\, ,
\\
& & \\
(d\chi)_{\underline{2}\underline{3}} 
& = &
\partial_{\underline{1}}H\, ,
\\
& & \\
(d\chi)_{\underline{3}\underline{1}} 
& = &
\partial_{\underline{2}}\left(W^{2}H\right)\, ,
\end{array}
\end{equation}

\noindent
whose integrability condition is

\begin{equation}
\label{eq:integrability}
\mathfrak{D}^{2}H\equiv \partial_{\underline{1}}\partial_{\underline{1}} H
+\partial_{\underline{2}}\partial_{\underline{2}}\left(W^{2} H\right)
+\partial_{\underline{3}}\partial_{\underline{3}} H 
= 0\, .
\end{equation}

As shown in Ref.~\cite{Chimento:2016run}, imposing different conditions on $W$
one can recover more restricted classes of metrics. In particular, when $W=1$
the 3-dimensional metric is flat and the constraint
Eqs.~(\ref{eq:constraintijcurved}) reduce to

\begin{equation} 
d\chi=\star_{3} dH\, ,
\end{equation} 

\noindent
which implies that $H$ is harmonic on $\mathbb{E}^{3}$ and the metric
Eq.~(\ref{eq:final_metric}) is a Gibbons-Hawking metric.

The curvature of these metrics has been computed in
Appendix~\ref{app-4dmetric} using the results of Appendix~\ref{app-3dmetric}
and we have also computed the curvature of the 5-dimensional metric
Eq.~(\ref{eq:conformastationaryd5}) for the above base space in
Appendix~\ref{app-5dmetric}. In what follows we use the frames defined in the
appendices.

The simplest non-trivial example of K\"ahler manifold admitting a holomorphic
isometry is the non-compact symmetric space
$\overline{\mathbb{CP}}^{2}=$SU$(1,2)/$U$(2)$ which is the base space of
AdS$_{5}$.\footnote{Actually, it is the only possible base space for AdS$_{5}$
  \cite{Gutowski:2004yv}.} Written in the conformastationary form
Eq.~(\ref{eq:conformastationaryd5}), AdS$_{5}$ is a U$(1)$ bundle over
$\overline{\mathbb{CP}}^{2}$ \cite{Gibbons:2011sg}, the non-compact version of
the Hopf fibrations studied in Ref.~\cite{Trautman:1977im}.  For the
convenience of the reader, we revisit this example in Appendix~\ref{app-ads5},
giving the functions $H$ and $W$ corresponding to $\overline{\mathbb{CP}}^{2}$
and describing how to rewrite this metric in more standard coordinates.

Assuming our base space is of the above form, then, we can continue our
analysis of the equations that determine the supersymmetric solutions of
minimal gauged supergravity.

To start with, if we choose a particular form for the complex structures
$\hat{\Phi}^{(2,3)}$ we can solve for $\hat{A}_{m}$ in Eqs.~(\ref{eq:df2=pf3})
and (\ref{eq:df3=-pf2}).

In the frame given by Eq.~\ref{eq:4dvierbein} and taking into account the
choice of $\hat{\Phi}^{(1)}$ already made in footnote~\ref{foot:JPhi}, we can
choose\footnote{The most general possible form for these matrices would be
  $\hat{\Phi}^{(2)\, \prime} =\cos{\theta}\, \hat{\Phi}^{(2)}+\sin{\theta}\,
  \hat{\Phi}^{(3)}$ and $ \hat{\Phi}^{(3)\, \prime} = \cos{\theta}\,
  \hat{\Phi}^{(3)} -\sin{\theta}\, \hat{\Phi}^{(2)}$, for some function
  $\theta$, in which case $\hat{A}\rightarrow \hat{A} -\tfrac{1}{g}d\theta$, which amounts
  to just a gauge transformation of the gauge fields.}

\begin{equation}
\label{eq:J23_matrices} 
(\hat{\Phi}^{(2)}{}_{mn})
=
\begin{pmatrix} 
i\sigma_{2} & \phantom{-i}0_{2\times 2}\\ 
\phantom{i}0_{2\times 2} & - i\sigma_{2} \\
\end{pmatrix}\, ,
\hspace{1.5cm}
(\hat{\Phi}^{(3)}{}_{mn})
=
\begin{pmatrix} 
\phantom{-i}0_{2\times 2} & -i\sigma_{2} \\ 
- i\sigma_{2} & \phantom{-i}0_{2\times 2} \\
\end{pmatrix}\, .
\end{equation} 

Then, we find that the flat components of $\hat{A}$ are given by 

\begin{equation} 
g\hat{A}_{\sharp} = -H^{-1/2}\overline{\omega}_{112}\, ,
\hspace{1.5cm}
g\hat{A}_{i} =
-H^{-1/2}\overline{\omega}_{i13}\, ,
\end{equation} 

\noindent
and, taking into account the 3-dimensional metric at hands, we find that we
can write all the components of $\hat{A}_{m}$ in the compact form

\begin{equation}
\label{eq:P_vector} 
g\hat{A}_{m} = \hat{J}_{m}{}^{n}\,\partial_{n}\log{W}\, ,
\end{equation}

\noindent
and, thus, we have solved the three Eqs.~(\ref{eq:df1=0})-(\ref{eq:df3=-pf2})
(or, equivalently, the original Eq.~(\ref{eq:dfAf})) in terms of the functions
that define the base space.

The consistency of this solution can be checked through the relation between
the field strength $\hat{F}_{mn}$ and the Ricci 2-form $\hat{\mathfrak{R}}_{mn}$
Eq.~(\ref{eq:R=dP}): using this relation, we get

\begin{eqnarray}
\hat{\mathfrak{R}}_{mn}
& = &
-g\hat{F}_{mn}
=
2\hat{\nabla}_{[m|} \hat{\nabla}_{p}\log{W}\hat{J}{}^{p}{}_{|n]}\, ,
\\
& & \nonumber \\
\label{eq:Rmn}
\hat{R}_{mn} 
& = & 
\hat{\nabla}_{m} \hat{\nabla}_{n}\log{W} 
+\hat{J}_{m}{}^{p}\hat{J}_{n}{}^{q}
\hat{\nabla}_{p} \hat{\nabla}_{q}\log{W}\, ,
\\
& & \nonumber \\
\hat{R} 
& = & 
\hat{\nabla}^{2}\log{W^{2}}\, .
\end{eqnarray}

These expressions can be compared with the direct computation of the Ricci
tensor and scalar in Appendix~\ref{app-4dmetric}. The expression of the Ricci
scalar can be used in Eq.~(\ref{eq:RicciVf}) to obtain a direct expression of
the metric function $\hat{f}$ in terms of the functions that define the base
space:

\begin{equation}
\label{eq:f-1}
\hat{f}^{-1} = \frac{1}{8g^{2}}\hat{\nabla}^{2}\log{W^{2}}\, .  
\end{equation}

Now Eq.~(\ref{eq:susy_{3}}) (or, equivalently, the original
Eq.~(\ref{eq:Fminus})) is also completely solved by Eqs.~(\ref{eq:P_vector})
and (\ref{eq:f-1}), and the only equations that remain to be solved are
Eqs.~(\ref{eq:susy_{2}}) and (\ref{eq:susy_4-2}). Observe that, since both
$\hat{f}^{-1}$ and $\hat{F}_{mn}$ are given by second-order derivatives, the
remaining equations will be, at most, of fourth order in derivatives, instead
of of sixth order as in Ref.~\cite{Gutowski:2004yv}.  We are going to try to
rewrite them in a simpler form as in the ungauged case.

Every (anti-)selfdual 2-form $\mathcal{F}^{\pm}$ on the four dimensional
K\"ahler base space can be written in terms of a 1-form living on the
3-dimensional space $\vartheta=\vartheta_{\underline{i}}dx^{i}$ as

\begin{equation}
\mathcal{F}^{\pm}
= 
e^{\sharp}\wedge \vartheta \pm \tfrac{1}{2} H\star_{3}\vartheta\, .
\end{equation}

The 2-forms we consider here are also $z$-independent and so will the
components of the corresponding 1-forms be. Thus, we introduce the
$z$-independent 3-dimensional 1-forms $\Lambda$, $\Sigma$, and $\Omega_{\pm}$
defined by

\begin{eqnarray}
\label{eq:simp_F+}
\hat{F}^{+}
& = &
-\tfrac{1}{2} \left( dz+\chi \right)\wedge \Lambda
-\tfrac{1}{2} H \star_{3}\Lambda\, ,
\\
& & \nonumber \\
\hat{F}^{-}
& = &
-\tfrac{1}{2} \left( dz+\chi \right)\wedge \Sigma
+\tfrac{1}{2} H \star_{3}\Sigma\, ,
\\
& & \nonumber \\
(d\hat{\omega})^{\pm}
& = &
\left(dz+\chi \right)\wedge \Omega^{\pm} \pm H \star_{3}\Omega^{\pm}\, ,
\end{eqnarray}

Comparing the expression of $\hat{F}^{-}$ with Eq.~(\ref{eq:susy_{3}}) and
those of $\hat{F}^{+}$ and $(d\omega)^{+}$ with Eq.~(\ref{eq:susy_{2}}) we
conclude that

\begin{eqnarray}
\label{eq:simp_Sigma_f}
\Sigma
& = & 
4g\hat{f}^{-1} dx^{2}\, ,
\\
& & \nonumber \\
\label{eq:Omega+}
\Omega^{+}
& = &
-\tfrac{\sqrt{3}}{4} \hat{f}^{-1} \Lambda\, .
\end{eqnarray}

Requiring the closure of $\hat{F}=\hat{F}^{+}+\hat{F}^{-}$ one gets

\begin{equation}
d\left( \Lambda +\Sigma \right)=0\, ,
\end{equation}

\noindent
which means that, locally, 

\begin{equation}
\label{eq:simp_Lambda_dW_Sigma}
\Lambda =d \left( K/H\right)-\Sigma\, ,
\end{equation}

\noindent
for some functions $K$.

From the same condition, using Eq.~(\ref{eq:integrability}) and the definition
of the operator $\mathfrak{D}^{2}$ in that equation, one also gets

\begin{equation}
\label{eq:KI_laplacian}
\mathfrak{D}^{2} K
=
8g\, \partial_{\underline{2}}\left( H W^{2}\hat{f}^{-1}\right)\, .
\end{equation}

Using Eq.~(\ref{eq:R=dP}) and the equations in the Appendices to compute the
Ricci 2-form for a metric of the kind we are considering here, one finds

\begin{equation}
\label{eq:sigma_k_comb}
g K = \partial_{\underline{2}}\log{W^{2}} + \kappa H\, ,
\end{equation}

\noindent
where $\kappa$ is an arbitrary constant that reflects the possibility of
adding to the solution of the inhomogeneous equation (\ref{eq:KI_laplacian})
solutions of the homogeneous equation. This expression for $K$, together with Eq.~(\ref{eq:f-1}), automatically
solves the second-order equation  Eq.~(\ref{eq:KI_laplacian}).   
It is convenient to rewrite $\hat{\omega}$ as

\begin{equation}
\hat{\omega} 
= 
\omega_{z} \left( dz+\chi \right)+\omega\, , 
\hspace{1cm}
\omega = \omega_{\underline{i}}dx^{i}\, ,
\end{equation}

\noindent
in terms of which 

\begin{equation}
\label{eq:omegapm}
\Omega^{\pm} 
= \pm\tfrac{1}{2}H^{-1}\left( \omega_{z} \star_{3} d\chi+\star_{3}  d\omega\right)
-\tfrac{1}{2} d\omega_{z}\, .
\end{equation}

\noindent
From Eqs.~(\ref{eq:Omega+}) and (\ref{eq:simp_Lambda_dW_Sigma}) we find that 

\begin{equation}
\Omega^{+} 
= 
-\tfrac{\sqrt{3}}{4}\hat{f}^{-1}
\left[d\left(K/H\right)-\Sigma\right]\, ,  
\end{equation}

\noindent
and, then, from Eq.~(\ref{eq:omegapm}), we find that 

\begin{equation}
\Omega^{-} 
= 
-\Omega^{+} -d\omega_{z}
=
\tfrac{\sqrt{3}}{4}\hat{f}^{-1}
\left[d\left(K/H\right)-\Sigma\right]-d\omega_{z}\, .
\end{equation}

\noindent
Using either of the last two equations in Eq.~(\ref{eq:omegapm}) one gets an
equation for $\omega$:

\begin{equation}
\label{eq:simp_dhatomega}
d\omega
=
H\star_{3}d\omega_{z}-\omega_{z} d\chi 
-\tfrac{\sqrt{3}}{2}\hat{f}^{-1}H\star_{3}
\left[d\left(K/H \right)-\Sigma \right]\, .
\end{equation}

Before calculating its integrability condition it is convenient to make a
change of variables (identical to the one made in the ungauged case) to
(partially) ``symplectic-diagonalize`` the right-hand side. Thus, we define
$L$ and $M$ through

\begin{equation}
\label{eq:newvariables}
\begin{array}{rcl}
\hat{f}^{-1}
& \equiv &
L + \tfrac{1}{12}K^{2}/H\, ,
\\    
& & \\
\omega_{z}
& \equiv & 
M +\tfrac{\sqrt{3}}{4} LK/H
+\tfrac{1}{24\sqrt{3}}K^{3}/H^{2}\, .
\end{array}
\end{equation}

Substituting these two expressions into Eq.~(\ref{eq:simp_dhatomega}) and
using the relation between the 1-form $\chi$ and the functions $H$ and $W$,
Eqs.~(\ref{eq:constraintijcurved}), the equation for $\omega$ takes the
form\footnote{We have left one $\omega_{z}$ in order to get a more compact
  expression.}

\begin{equation}
d\omega
=
\star_{3}
\left\{
HdM -MdH +\tfrac{\sqrt{3}}{4} \left(KdL-LdK \right)
-H\left(\omega_{z}\partial_{\underline{2}}\log{W^{2}} 
-2\sqrt{3}g \hat{f}^{-2} \right)dx^{2}
\right\}\, ,  
\end{equation}

\noindent
and its integrability equation is just\footnote{One has $\star_{3}d\star_{3}d =
  \overline{\nabla}^{2}$.}

\begin{equation}
\label{eq:integrabilityomegahat}
  \begin{array}{rcl}
H\overline{\nabla}^{2}M -M\overline{\nabla}^{2}H 
+\tfrac{\sqrt{3}}{4} \left(K\overline{\nabla}^{2}L
-L\overline{\nabla}^{2}K \right)\hspace{2cm}
& & \\
& & \\
-{\displaystyle\frac{1}{W^{2}}}
\partial_{\underline{2}}\left\{
HW^{2}\left(\omega_{z}\partial_{\underline{2}}\log{W^{2}} 
-2\sqrt{3}g \hat{f}^{-2} \right)\right\}
& = & 0\, .  
\end{array}
\end{equation}

This equation can be simplified by using the equations satisfied by the
functions $H$ and $K$ (\ref{eq:integrability}) and (\ref{eq:KI_laplacian}),
respectively. We postpone doing this until we derive the equation for $L$,
which follows from Eq.~(\ref{eq:susy_4-2}). First of all, observe that, with
our choice of complex structure Eq.~(\ref{eq:J_matrix})

\begin{equation}
\hat{J}\cdot (d\hat{\omega}) 
= 
4(d\hat{\omega})^{-}_{02} 
= 
4 \Omega^{-}_{\underline{2}}  
=
\sqrt{3}\hat{f}^{-1}
\left[\partial_{\underline{2}}\left(K/H\right)-4g\hat{f}^{-1}\right]
-\partial_{\underline{2}}\omega_{z}\, .
\end{equation}

\noindent
On the other hand, we have

\begin{equation}
  \begin{array}{rcl}
\hat{\nabla}^{2}\hat{f}^{-1}
& = &
H^{-1}
\overline{\nabla}^{2}\hat{f}^{-1}\, ,
\\
& & \\
\hat{F}\cdot\hat{\star}\hat{F}
 & = &
{\displaystyle
\Lambda_{m}\Lambda_{m}  
-\Sigma_{m}\Sigma_{m}
=
\partial_{m}(K/H)  \partial_{m}(K/H)
-2\Sigma_{m} \partial_{m}(K/H)\, ,
}
\\
& & \\
{\displaystyle
H\partial_{m}(K/H)\partial_{m}(K/H)
}
& = & 
{\displaystyle
\overline{\nabla}^{2} \left(\frac{K^{2}}{2H}\right)
+\frac{K^{2}}{2H^{2}}\overline{\nabla}^{2}H
-\frac{K\overline{\nabla}^{2}K}{H}\, ,
}
\\
\end{array}
\end{equation}

\noindent
and, using all these partial results into Eq.~(\ref{eq:susy_4-2}), and (not
everywhere, for the sake of simplicity) the new variables
Eqs.~(\ref{eq:newvariables}), we arrive at

\begin{equation}
\label{eq:simp_lapl_hI_1}
\begin{array}{rcl}
\overline{\nabla}^{2} L 
-\tfrac{1}{12}(K/H)^{2}\overline{\nabla}^{2}H  
+\tfrac{1}{6}(K/H)\overline{\nabla}^{2}K
+\tfrac{7}{3}gH\hat{f}^{-1} \partial_{\underline{2}}(K/H)
& & 
\\ 
& & \\
-\tfrac{4}{\sqrt{3}}gH\partial_{\underline{2}}\omega_{z} 
-4g^{2}H\hat{f}^{-2}
& = &
0\, .
\end{array}
\end{equation}

We can now use the relation between the 3-dimensional Laplacian and the
$\mathfrak{D}^{2}$ operator and the equations for the 
functions $H$ and $K$ (\ref{eq:integrability}) and
(\ref{eq:KI_laplacian})

\begin{equation}
  \begin{array}{rcl}
\overline{\nabla}^{2}H  
& = &
{\displaystyle
\frac{\mathfrak{D}^{2}H}{W^{2}}
-\partial_{\underline{2}}H\frac{\partial_{\underline{2}}W^{2}}{W^{2}}
-H\frac{\partial^{2}_{\underline{2}}W^{2}}{W^{2}}
}
=
{\displaystyle
-\partial_{\underline{2}}H\frac{\partial_{\underline{2}}W^{2}}{W^{2}}
-H\frac{\partial^{2}_{\underline{2}}W^{2}}{W^{2}}
}
\, ,
\\
& & \\
\overline{\nabla}^{2}K
& = &
{\displaystyle
\frac{\mathfrak{D}^{2}K}{W^{2}}
-\partial_{\underline{2}}K\frac{\partial_{\underline{2}}W^{2}}{W^{2}}
-K\frac{\partial^{2}_{\underline{2}}W^{2}}{W^{2}}
}
=
{\displaystyle
\frac{8g}{W^{2}} \partial_{\underline{2}}(HW^{2}\hat{f}^{-1})
-\partial_{\underline{2}}K\frac{\partial_{\underline{2}}W^{2}}{W^{2}}
-K\frac{\partial^{2}_{\underline{2}}W^{2}}{W^{2}}
}
\, ,
\\
\end{array}
\end{equation}

\noindent
and, setting $\kappa=0$ for simplicity from now on, the equation for $L$
becomes

\begin{equation}
\label{eq:pure_maxwell_LI}
\overline{\nabla}^{2}L
=
4 H \left( g L \right)^{2}-\tfrac{2}{3} L\left( g K \right)^{2}-\tfrac{4}{3} gL
 \partial_{\underline{2}}K
-\tfrac{1}{3} g K \partial_{\underline{2}}L 
+ \tfrac{4}{\sqrt{3}}H g \partial_{\underline{2}}M\, .
\end{equation}

\noindent
Using this equation in the integrability condition for the $\omega$ equation,
Eq.~(\ref{eq:integrabilityomegahat})  we get 

\begin{equation}
\label{eq:pure_integrabilityomegahat3}
\overline{\nabla}^{2}M
=
-\sqrt{3}g L\left( g K L+2 \partial_{\underline{2}}L \right)\, .
\end{equation}

While the appearance of these equations is quite compact, we have to take into
account that the functions appearing in them are not totally independent.
Using Eqs.~(\ref{eq:integrability}),(\ref{eq:f-1}),(\ref{eq:sigma_k_comb}) and
(\ref{eq:newvariables}) we find the following equations that have to be added
to these:

\begin{eqnarray}
\label{eq:integrabilityagain}
\mathfrak{D}^{2}H
& = & 
0\, ,
\\
& & \nonumber \\
\label{eq:gK}
g K 
& = &
\partial_{\underline{2}}\log{W}^{2} \, ,  
\\
& & \nonumber \\
\label{eq:L}
L
& = & 
\frac{1}{8g^{2}H}
\left\{
\overline{\nabla}^{2}\log{W^{2}}
-\tfrac{2}{3}\left(\partial_{\underline{2}}\log{W}^{2}\right)^{2}
\right\}\, .
\end{eqnarray}

Substituting them in the other two, we get fourth order differential equations
for $H,M,W$.

As was first noted in Ref.~\cite{Figueras:2006xx} not every K\"ahler base
space can give rise to a supersymmetric solution. This can be seen here as
follows: multiplying Eq.~(\ref{eq:pure_integrabilityomegahat3}) by $W^2$,
differentiating with respect to $x^2$, eliminating $\partial_{\underline{2}}M$
from the resulting equation with Eq.~(\ref{eq:pure_maxwell_LI}), and using
Eqs.~(\ref{eq:gK}) and (\ref{eq:L}) one gets a sixth order differential
equation involving only $H$ and $W^{2}$, which are the functions that
determine the K\"ahler base space. This is then a constraint on the admissible
base spaces, and while we did not check this explicitly it is likely to be
equivalent to the constraint found in Ref.~\cite{Cassani:2015upa} for an
arbitrary K\"ahler base space.

\section{Examples}
\label{sec-examples}

Before we set out to solve the equations, in order to gain some insight, it is
convenient to rewrite some simple and well-known supersymmetric solutions in
the form we are proposing here.

\subsection{Reissner-Nordstr\"om-AdS$_{5}$}
\label{sec-RNADS5example}

Thus, let us consider the asymptotically AdS$_{5}$ Reissner-Nordstr\"om
(RN-AdS$_{5}$) solutions, which are given by the metric and vector
field

\begin{equation}
\begin{array}{rcl}
ds^{2}
& = &
[k+h(r)+\tfrac{1}{3}g^{2}r^{2}]dt^{2} 
-{\displaystyle\frac{dr^{2}}{[k+h(r)+\tfrac{1}{3}g^{2}r^{2}]}}
-r^{2}d\Omega^{2}_{(3,k)}\, ,
\\
& & \\ 
A
& = &
{\displaystyle\frac{3q}{r^{2}}}dt\, ,   
\\
& & \\ 
h(r)
& = & 
{\displaystyle
-\frac{2M}{r^{2}}+\frac{3q^{2}}{r^{4}} 
}
\end{array}
\end{equation}

\noindent
where $M$ is the mass, $q$, the electric charge that we will assume to be
positive for the sake of simplicity,\footnote{The mass and charge are defined
  in units in which
\begin{equation}
\label{eq:units1}
\frac{16\pi G^{(5)}_{N}}{3\omega_{(3)}}= 2\, ,    
\end{equation}
where $G^{(5)}_{N}$ is the 5-dimensional Newton constant and $\omega_{(3)}$
the volume of the round 3-sphere of unit radius, for the $k=1$ case.
Equivalently, we have chosen units such that
\begin{equation}
\label{eq:units2}
\frac{3\pi}{4G_{N}^{(5)}}=1\, .   
\end{equation}
}
$k=1,0,-1$ the curvature of the 3-dimensional metric
$d\Omega^{2}_{(3,k)}$. More explicitly, for $k=1$ $d\Omega^{2}_{(3,1)}\equiv
d\Omega^{2}_{(3)} $ is the metric of the round sphere of unit radius

\begin{equation}
d\Omega^{2}_{(3)}
= 
\tfrac{1}{4}  
\left[
(d\psi+\cos{\theta}\, d\varphi)^{2}
+
d\Omega^{2}_{(2)}
\right]\, ,
\hspace{1cm}
d\Omega^{2}_{(2)}
= 
d\theta^{2} +\sin^{2}{\theta}\, d\varphi^{2}\, ,
\end{equation}

\noindent
for $k=0$ $d\Omega^{2}_{(3,0)}$ is the metric of $\mathbb{E}^{3}$ with the
normalization

\begin{equation}
d\Omega^{2}_{(3,0)} 
= 
\tfrac{1}{4}  
\left[
d\psi^{2}
+
d\Omega^{2}_{(2,0)}
\right]\, , 
\hspace{1cm}
d\Omega^{2}_{(2,0)}
=
(dx^{1})^{2} +(dx^{3})^{2}\, ,
\end{equation}

\noindent
and for $k=-1$ $d\Omega^{2}_{(3,-1)}$ is the metric of $\mathbb{H}_{3}$. We
have not succeeded in writing this metric in the form of a fibration over
another 2-dimensional space and, therefore, we will not be able to rewrite the
corresponding solution in the form required by supersymmetry. Actually, it is
well known that supersymmetry requires the following relation between the
mass, the charge and $k$:\footnote{The supersymmetric $k=1$ RN-AdS$_{5}$
  solution was first found in Ref.~\cite{London:1995ib}. In
  Ref.~\cite{Cassani:2015upa} it was shown that it is the only supersymmetric
  solution with $\mathbb{R}\times$SO$(4)$ isometry group. Here , we present it
  in the canoncial supersymmetric form. The $k=0,-1$ cases have isometry
  groups $\mathbb{R}\times$ISO$(3)$ and $\mathbb{R}\times$SO$(2,2)$,
  respectively.}

\begin{equation}
\label {eq:susyrel}
M^{2}= 3kq^{2}\, ,  
\end{equation}

\noindent
and, therefore, we do not expect supersymmetric solutions for $k=-1$, except
pure AdS$_{5}$ space. However, pure AdS$_{5}$ space cannot be described in the
form required by $k=-1$.

We are going to rederive this result by rewriting the metric in the canonical
form Eqs.~(\ref{eq:conformastationaryd5}), (\ref{eq:final_metric}) and
(\ref{eq:constraintijcurved}) we are proposing, identifying the functions
$\hat{f},H,W,\omega_{z}$ and the 1-forms $\chi,\omega$ and checking that they
satisfy the equations that we have derived from supersymmetry.

First, we transform the coordinate $\psi=z-\frac{2}{\sqrt{3}}gt$ and perform a
gauge transformation of the vector field to get

\begin{equation}
\begin{array}{rcl}
ds^{2}
& = &
(k+h)
\left[ 
dt +\tfrac{1}{2\sqrt{3}}g {\displaystyle\frac{r^{2}}{k+h}} (dz+\chi_{(k)})    
\right]^{2}
-{\displaystyle\frac{r^{2}}{4(k+h)}}
[k+h(r)+\tfrac{1}{3}g^{2}r^{2}](dz+\chi_{(k)})^{2}
\\
& & \\
& & 
-{\displaystyle\frac{dr^{2}}{[k+h(r)+\tfrac{1}{3}g^{2}r^{2}]}}
-\frac{1}{4}r^{2}\, d\Omega^{2}_{(2,k)}\, ,
\\
& & \\
A
& = & 
-\sqrt{3}\left(\delta-{\displaystyle\frac{\sqrt{3}q}{r^{2}}}\right)
\left[ 
dt +\tfrac{1}{2\sqrt{3}}g {\displaystyle\frac{r^{2}}{k+h}} (dz+\chi_{(k)})    
\right]
+\tfrac{1}{2}g{\displaystyle\frac{\delta r^{2}-\sqrt{3}q}{k+h}} (dz+\chi_{(k)})\, ,
\end{array}
\end{equation}

\noindent
where $\chi_{(1)}=\cos{\theta}d\varphi$, $\chi_{(0)}=0$ and $\delta$ is an
arbitrary constant. Observe that, for $h=0$ (pure AdS$_{5}$) this
transformation can only be made for $k=1$. We will have to study more
carefully the asymptotic behaviour of the transformed solution for $k=0$.

We also need to rewrite $d\Omega^{2}_{(2,k)}$ and, correspondingly
$\chi_{(k)}$ as in Eq.~(\ref{eq:Phidef}).  $\hat{f},\omega_{z},\chi,H$ and $W$
can be read immediately from $g_{tt},g_{tz},g_{zz}$ and
$g_{\underline{1}\underline{1}} = g_{\underline{3}\underline{3}}$,
respectively. $g_{rr}$ should be given by $\hat{f}^{-1}H$, but this only
happens after a change of coordinates $r=2\varrho^{1/2}$.\footnote{We denote
  by $\varrho$ the coordinate $x^{2}$ in Eq.~(\ref{eq:final_metric}).} The
final result is

\begin{equation}
\begin{array}{rcl}
\hat{f} 
& = & 
[k+h(\varrho)]^{1/2}\, ,
\\
& & \\
H
& = &
{\displaystyle
\frac{[k+h(\varrho)]^{1/2}}{\varrho[k+h(\varrho)+\tfrac{4}{3}g^{2}\varrho]}
}\, ,
\\
& & \\
W^{2}
& = & 
\varrho^{2} [k+h(\varrho)+\tfrac{4}{3}g^{2}\varrho] \Phi(x^{1},x^{3})\, ,
\\
& & \\
\omega_{z}
& = & 
\tfrac{2}{\sqrt{3}}g {\displaystyle \frac{\varrho}{k+h(\varrho)}}\, ,
\\
& & \\
\chi
& = & 
\chi_{(k)}\, ,
\\
& & \\
\omega
& = & 
0\, ,
\\
& & \\
A
& = & 
-\sqrt{3}\left(\delta
-{\displaystyle\frac{\sqrt{3}q}{4\varrho}}\right)[dt+\omega_{z}(dz+\chi_{(k)})]
+\tfrac{1}{2}g{\displaystyle\frac{4\delta\varrho -\sqrt{3}q}{k+h}} (dz+\chi_{(k)})\, ,
\end{array}
\end{equation}

\noindent
where, now

\begin{equation}
h(\varrho) = -\frac{M}{2\varrho}+ \frac{3q^{2}}{16\varrho^{2}}\, ,
\end{equation}

\noindent
and $\Phi_{(k)}(x^{1},x^{3})$ and $\chi_{(k)}$ have been defined in
Eqs.~(\ref{eq:Phidef}).

Eq.~(\ref{eq:integrabilityagain}) is satisfied if 

\begin{equation}
k+h = \left(k-\frac{\sqrt{3}q}{4\varrho} \right)^{2}\, ,   
\end{equation}

\noindent
which implies the supersymmetry relation Eq.~(\ref{eq:susyrel}).

The 1-form potential coincides with the one in
Eq.~(\ref{eq:completevectorfields}) if $\delta =k$. In particular, and Eq.~(\ref{eq:P_vector})
is satisfied up to a gauge transformation.

The rest of the equations are also satisfied. 

To summarize, the supersymmetric RN-AdS$_{5}$ solutions for $k=0,1$ are given
by 

\begin{equation}
\begin{array}{rcl}
\hat{f} 
& = & 
k-{\displaystyle\frac{\sqrt{3}q}{4\varrho}}\, ,
\\
& & \\
H
& = &
{\displaystyle
\frac{k\varrho-\tfrac{\sqrt{3}}{4}q}{\tfrac{4}{3}g^{2}\varrho^{3} +k^{2}\varrho^{2}
  -\frac{\sqrt{3}}{2}q\varrho +\tfrac{3}{16}q^{2}}
}\, ,
\\
& & \\
W^{2}
& = & 
[\tfrac{4}{3}g^{2}\varrho^{3} +k^{2}\varrho^{2}
  -\frac{\sqrt{3}}{2}q\varrho +\tfrac{3}{16}q^{2}] \Phi(x^{1},x^{3})\, ,
\\
& & \\
\omega_{z}
& = & 
\tfrac{2}{\sqrt{3}}g {\displaystyle \frac{\varrho^{3}}{k^{2}\varrho^{2}
  -\frac{\sqrt{3}}{2}q\varrho +\tfrac{3}{16}q^{2}}}\, ,
\\
& & \\
\chi
& = & 
\chi_{(k)}\, ,
\\
& & \\
\omega
& = & 
0\, ,
\\
& & \\
A
& = & 
-\sqrt{3}\hat{f}[dt+\omega_{z}(dz+\chi_{(k)})]
+2g\varrho \hat{f}^{-1} (dz+\chi_{(k)})\, .
\end{array}
\end{equation}

Setting $q=0$ in the  $k=1$ case we get AdS$_{5}$ written in the
canonical supersymmetric form

\begin{equation}
\label{eq:ads5k1}
ds^{2} 
=
\left[ 
dt +\tfrac{2}{\sqrt{3}}g \varrho (dz+\cos{\theta}d\varphi)    
\right]^{2}
-\varrho [1+\tfrac{4}{3}g^{2}\varrho](dz+\cos{\theta}d\varphi)^{2}
-{\displaystyle\frac{d\varrho^{2}}{\varrho[1+\tfrac{4}{3}g^{2}\varrho]}}
-\varrho\, d\Omega^{2}_{(2)}\, .
\end{equation}

In Appendix~\ref{app-ads5} we discuss the relation between this form of
AdS$_{5}$ and more popular forms of the same metric with $g=\sqrt{3}$. As it
is shown there, the base space is the symmetric K\"ahler space
$\overline{\mathbb{CP}}^{2}$. In Ref.~\cite{Gutowski:2004ez} it has been shown
that this is the only possible base space for AdS$_{5}$. However,
$\overline{\mathbb{CP}}^{2}$ can be written in different ways, using the
metric of S$^{2}$, $\mathbb{E}^{2}$ or $\mathbb{H}_{2}$, and we are going to
see in the next example that there are 3 associated canonical metrics for
AdS$_{5}$ that can be used to construct more general solutions. The
construction of these metrics is explained in Appendix~\ref{app-ads5}.

In the $k=0$ case we also get AdS$_{5}$, but in different (non-canonical)
coordinates:

\begin{equation}
ds^{2} 
=
\varrho \left[ 
\tfrac{4}{\sqrt{3}}gdt dz + (dx^{1})^{2}+(dx^{3})^{2}
\right]
-{\displaystyle\frac{d\varrho^{2}}{\tfrac{4}{3}g^{2}\varrho^{2}}}\, .
\end{equation}

The $\varrho\rightarrow \infty$ limit is, in these solutions, equivalent to
setting $q=0$. In the $\varrho\rightarrow 0$ limits both solutions give
the following singular geometries

\begin{equation}
  \begin{array}{rcl}
ds^{2}
& = &
{\displaystyle
\frac{3q^{2}}{4\varrho^{2}} dt^{2} 
-\varrho   
\left[\frac{16}{3q^{2}}d\varrho^{2} +4d\Omega^{2}_{(3)}
\right]\, ,
}
\\
& & \\
ds^{2}
& = &
{\displaystyle
\frac{3q^{2}}{4\varrho^{2}} dt^{2} 
-\varrho   
\left[\frac{16}{3q^{2}}d\varrho^{2} +dz^{2} + (dx^{1})^{2} + (dx^{3})^{2}
\right]\, ,
}
\end{array}
\end{equation}

\noindent
which are also examples of supersymmetric solutions written in the
canonical form.

Finally, in the $k=1$ case, the supersymmetric Killing vector becomes null at
$\varrho = \tfrac{\sqrt{3}}{4}q$, indicating the possible existence of a
Killing horizon which would also be a candidate to event horizon. It is
convenient to work with the shifted coordinate $\varrho^{\prime}=
\varrho-\tfrac{\sqrt{3}}{4}q$, which is zero at the point of interest. The
radial coordinate of the
solutions that we are going to present in the next section also vanishes at
the same point and, in order to ease the comparison between the solutions, we
rewrite here the $k=1$ RN-AdS$_{5}$ solution in the shifted radial coordinate
(suppressing the primes):

\begin{equation}
\label{eq:RNAdS5k=1}
\begin{array}{rcl}
\hat{f} 
& = & 
\varrho\, \left(\varrho+\tfrac{\sqrt{3}}{4}q\right)^{-1}\, ,
\\
& & \\
H
& = &
{\displaystyle
\varrho
\left[\frac{4g^{2}}{3}\varrho^{3} +(1+\sqrt{3}g^{2}q)\varrho^{2}
  +\frac{3g^{2}q^{2}}{4}\varrho +\frac{\sqrt{3}g^{2}q^{3}}{16}
\right]^{-1}
}\, ,
\\
& & \\
W^{2}H
& = & 
\varrho\, \Phi(x^{1},x^{3})\, ,
\\
& & \\
\omega_{z}
& = & 
{\displaystyle
\frac{2g}{\sqrt{3}}
\varrho^{-2}\, \left(\varrho+\frac{\sqrt{3}q}{4} \right)^{3}\, ,
}
\end{array}
\end{equation}

\noindent
and, in the $\varrho\rightarrow 0$ limit, the metric takes the form

\begin{equation}
ds^{2}
=
\frac{16}{3q^{2}}\varrho^{2}dt^{2} +gqdt(dz+\chi_{(1)})
-\frac{4}{g^{2}q^{2}}d\varrho^{2} -\sqrt{3}qd\Omega^{2}_{(3)}\, ,
\end{equation}

\noindent
and does not coincide with any of the near-horizon metrics constructed in
Ref.~\cite{Gutowski:2004yv}. In particular, observe that the metric of the
hypersurface $\varrho=0$ has rank four, which means that it cannot be null. 
It is a well-known fact that the supersymmetric RN-AdS$_{5}$ solution has a
naked singularity.

\subsection{AdS$_{5}$}
\label{sec-ADS5example}

We have found three interesting ways of writing AdS$_{5}$ in the
supersymmetric canonical form:

\begin{equation}
\label{eq:ads5k}
ds^{2}
=\left[ dt+\tfrac{2}{\sqrt{3}}g \varrho  (dz+\chi_{(k)})\right]^{2}
-\varrho(k+\tfrac{4}{3}g^{2}\varrho)(dz+\chi_{(k)})^{2}
-\frac{d\varrho^{2}}{\varrho(k+\tfrac{4}{3}g^{2}\varrho)}
-\varrho d\Omega_{(2,k)}^{2}\, ,
\end{equation}

\noindent
where $d\Omega_{(2,k)}^{2}$ (the metric of the unit 2-sphere, plane and
hyperplane for, respectively, $k=1,0$ and $-1$) and $\chi_{(k)}$ are given by
Eqs.~(\ref{eq:Phidef}). The case $k=1$ has been given in
Eq.~(\ref{eq:ads5k1}). The base space has a metric of the form
Eq.~(\ref{eq:final_metric}) with

\begin{equation}
H^{-1}=\varrho(k+\tfrac{4}{3}g^{2}\varrho)\, ,
\hspace{1cm}
W^{2}H=\varrho \Phi_{(k)}(x^{1},x^{3})\,   
\hspace{1cm}
\chi=\chi_{(k)}\, .
\end{equation}

\noindent
It is, by construction, a K\"ahler space with one holomorphic isometry. In
agreement with Ref.~\cite{Gutowski:2004ez}, this metric is that of
$\overline{\mathbb{CP}}^{2}$ for the three values of $k$ as shown in
Appendix~\ref{app-ads5}. In the full 5-dimensional metric, the coordinate $z$
has a different causal character in each case: the norm of the Killing vector
$\partial_{z}$ is $g_{zz}=-k\varrho$, and, since $\varrho$ has to be positive
in all cases, the coordinate $z$ turns out to be null for $k=0$ and timelike
for $k=-1$.

For $k=1$, as shown in Appendix~\ref{app-ads5}, we can go to an unrotating
coordinate system with the change $z=\psi+\tfrac{2}{\sqrt{3}}gt$

\begin{equation}
\label{eq:ads5k1unrotating}
ds^{2}
=
(1+\tfrac{4}{3}g^{2}\varrho) dt^{2}
-\varrho (d\psi+\chi_{(1)})^{2}
-\frac{d\varrho^{2}}{\varrho(1+\tfrac{4}{3}g^{2}\varrho)}
-\varrho d\Omega_{(2,1)}^{2}\, ,
\end{equation}

\noindent
which is well defined for all positive values of $\varrho$.  For $k=-1$,
changing $z=\psi-\tfrac{2}{\sqrt{3}}gt$ we get

\begin{equation}
\label{eq:ads5k-1unrotating}
ds^{2} 
= 
\varrho (d\psi +\chi_{(-1)})^{2}
-(\tfrac{4}{3}g^{2}\varrho -1)dt^{2} 
-\frac{d\varrho^{2}}{\varrho(\tfrac{4}{3}g^{2}\varrho -1)}
-\varrho d\Omega_{(2,-1)}^{2}\, ,
\end{equation}

\noindent
which is well defined for $\varrho > \frac{3}{4 g^{2}}$ and shows the timelike
character of $\psi$ and the spacelike character of $t$. For $k=0$ there is no
analogous transformation.  It is worth stressing that the $\varrho$
coordinates of these three AdS$_{5}$ metrics are different as the $z$ and $t$
coordinates are.

In the next section we are going to find two families of solutions using
ansatzs adapted to these three forms of AdS$_{5}$. Only by using them the
equations of motion become tractable. Actually, rewriting the solutions found
using the $k=0,-1$ ansatzs in the $k=1$ coordinates (more conventional and
better understood) although possible, leads to very complicated metrics. Thus, it
would be rather convenient to be able to analyze the asymptotic behaviour of
the $k=0,-1$ solutions and compute their conserved charges directly in the
$k=0,-1$ coordinates. 

Indeed, naively, some of the $k=0,-1$ solutions we are going to present seem
to approach the above $k=0,-1$ forms of the AdS$_{5}$ metric or the naive
asymptotic limit of those metrics. However, given the many subtleties that
arise in the study of asymptotically-AdS solutions, a more rigorous analysis
using Penrose's conformal techniques \cite{Penrose:1965am}, as in
Ref.~\cite{Ashtekar:1999jx} is required.

Let us first study the three AdS$_{5}$ metrics since, as we just discussed,
their asymptotic limits appear in the asymptotic limits of the most general
solutions. 

In the $k=1$ case the only spacelike coordinate which is not compact is
$\varrho\in [0,+\infty)$ and some components of the metric diverge in the
$\varrho\rightarrow +\infty$ limit. Thus, we make the coordinate
transformation $\varrho \equiv \tan^{2}{\xi}$, which brings the metric
Eq.~(\ref{eq:ads5k1unrotating}) to the form 

\begin{equation}
ds^{2}
=
4\cos^{-2}{\xi}\, \tilde{ds}{}^{2}\, ,  
\end{equation}

\noindent
where

\begin{equation}
\tilde{ds}{}^{2}
=
\left(\tfrac{1}{4}\cos^{2}{\xi} +(g/\sqrt{3})^{2}\sin^{2}{\xi} \right)dt^{2}
-\frac{d\xi^{2}}{4\left(\tfrac{1}{4}\cos^{2}{\xi} 
+(g/\sqrt{3})^{2}\sin^{2}{\xi} \right)}
-\sin^{2}{\xi}\, d\Omega^{2}_{(3)}\, .  
\end{equation}

\noindent
$\tilde{ds}{}^{2}$ is regular at $\xi=\pi/2$ ($\varrho\rightarrow +\infty$) and
becomes, at that point 

\begin{equation}
\tilde{ds}{}^{2}(\xi=\pi/2)
=
(g/\sqrt{3})^{2}dt^{2}-\tfrac{1}{4}d\Omega^{2}_{(3)}\, .   
\end{equation}

\noindent
This space is just $\mathbb{R}\times S^{3}$, whose conformal isometry group is
SO$(2,4)$. Since the conformal factor relating the metrics $\Omega=\cos{\xi}$
vanishes on the boundary $\xi=\pi/2$ but $\nabla_{a}\Omega$ does not,
according to the definition of Ref.~\cite{Ashtekar:1999jx}, the AdS$_{5}$
metric, in the $k=1$ form is asymptotically AdS$_{5}$ in the direction
$\rho\rightarrow \infty$\footnote{The Weyl tensor of AdS$_{5}$ of course vanishes identically.}.

In the $k=-1$ case there are two non-compact spacelike coordinates: $\varrho$
and $\theta$. We make in the metric Eq.~(\ref{eq:ads5k-1unrotating}) the
following changes of coordinates:

\begin{equation}
\varrho \equiv \tan^{2}{\xi}\, ,
\hspace{.5cm}
\psi \equiv \alpha+\beta\, , 
\hspace{.5cm}
\varphi=\alpha-\beta\, ,
\hspace{.5cm}
\sinh{(\theta/2)}=\tan{\eta}\, ,
\end{equation}

\noindent
finding 

\begin{equation}
ds^{2}
=
4\cos^{-2}{\xi}\cos^{-2}{\eta}\, \tilde{ds}{}^{2}\, ,  
\end{equation}

\noindent
with

\begin{align}
\tilde{ds}{}^{2}
 & =
\sin^{2}{\xi}\left[ d\alpha^{2} -\eta^{2}-\sin^{2}{\eta}d\beta^{2} \right]
\nonumber \\
& \nonumber  \\
&
-\cos^{2}{\eta}
\left[
\left( (g/\sqrt{3})^{2}\sin^{2}{\xi} -\tfrac{1}{4}\cos^{2}{\xi}\right)dt^{2}
+\frac{d\xi^{2}}{4\left( (g/\sqrt{3})^{2}\sin^{2}{\xi} -\tfrac{1}{4}\cos^{2}{\xi}\right)}
\right]\, .
\end{align}

In these coordinates, the boundary lies where the conformal factor
$\Omega=\cos{\xi}\cos{\eta}$ vanishes and it seems to correspond to two
different pieces: $\xi=\pi/2$ and $\eta=\pi/2$. The first piece has the
induced metric

\begin{equation}
\tilde{ds}^{2}(\xi=\pi/2)
=
d\alpha^{2}
-[d\eta^{2} 
+\sin^{2}{\eta}\, d\beta^{2}  
+\cos^{2}{\eta}\,d(gt/\sqrt{3})^{2}]\, ,
\end{equation}

\noindent
which is the metric of $\mathbb{R}\times S^{3}$, but now it is $\alpha$ the
coordinate that plays the r\^ole of time while $t$ is an angle. 

The second piece, though, has a singular metric. To understand the reason for
the existence of an apparent second piece of the boundary we can look at the
relation between the $\theta,\varrho$ coordinates of the $k=1$ and $k=-1$ case
since, in the $k=1$ case the boundary coincides exactly with the
$\varrho\rightarrow \infty$ limit. This relation can be inferred from the
$k=1$ and $k=-1$ parametrizations of $\overline{\mathbb{CP}^{2}}$ in
Appendix~\ref{app-ads5} and takes the form 

\begin{equation}
\varrho 
= 
\bar{\varrho}\cosh{\bar{\theta}/2} -\frac{3}{4g^{2}}\, ,
\hspace{1cm}
\tan{\theta/2} 
=
\frac{\frac{2}{\sqrt{3}}g\bar{\varrho}}{\sqrt{\frac{4}{3}g^{2}\bar{\varrho}-1}}
\sinh{\bar{\theta}/2}\, ,
\end{equation}

\noindent
where the barred coordinates correspond to the $k=-1$ case. While
$\bar{\varrho}\rightarrow \infty$ limit covers the same region as the
$\varrho\rightarrow \infty$ limit (the boundary), a subspace of the same
$\varrho\rightarrow \infty$ region with $\theta=\pm \pi$ can be also reached
in the limits $\bar{\theta}\rightarrow \pm \infty$. This subspace is covered
twice. We could, then, ignore the $\eta=\pi/2$ $\xi\neq \pi/2$ piece of the
boundary and consider just the $\xi =\pi/2$ one. However, the derivative of
the conformal factor vanishes on the boundary at $\eta=\pi/2$. We could
exclude these points to avoid this problem and add the second piece of the
boundary, but, as we have seen, the induced metric is not regular there. Thus,
at best, in the $k=-1$ coordinates we can only describe part of the boundary
and the solutions that use these coordinates asymptotically will have the same
problem.

Things are much more complicated in the $k=0$ case. It is convenient to
proceed in two steps. First, we redefine the $\varrho$ coordinate as in the
preceding cases in terms of $\xi$ and set $\xi=\pi/2$, getting

\begin{equation}
\label{eq:firstconformalmetrick0}
\tilde{ds}^{2}(\xi=\pi/2)
=
(g/\sqrt{3}) dt [dz+2(ydx-xdy)] 
-dx^{2}-dy^{2}\, . 
\end{equation}

\noindent
Then, we redefine $x=\tfrac{\zeta}{2}\cos{\varphi}$ and $y=\tfrac{\zeta}{2}\sin{\varphi}$, getting

\begin{equation}
\tilde{ds}^{2}(\xi=\pi/2)
=
(g/\sqrt{3}) dt [dz-\tfrac{\zeta^{2}}{2}d\varphi] 
-\tfrac14[d\zeta^{2}+\zeta^{2}d\varphi^{2}]\, ,
\end{equation}

\noindent
and shift the $t$ coordinate

\begin{equation}
\label{eq:tshift}
dt\rightarrow dt -4\frac{(2+\zeta^{2})dz -2 z \zeta d\zeta}{16z^{2}+(2+\zeta^{2})^{2}}\,,
\end{equation}

\noindent
which can be done since the added part is a closed 1-form. Finally, we make
the coordinate transformation\footnote{To obtain these coordinate changes one
  can consider the embedding of AdS$_{5}$ in $\mathbb{C}^{1,2}$ in terms of
  complex coordinates $Z^{0},Z^{1}, Z^{2}$. The ``correct'' limit giving the
  asymptotic boundary of AdS is the one obtained by sending $|Z^{0}|$ to
  infinity while leaving $Z^{1}$, $Z^{2}$ and the phase of $Z^{0}$
  independent. For any fixed value of $|Z^{0}|$, $Z^{1}$ and $Z^{2}$
  parametrize a 3-sphere. One then wants to choose coordinates such that, at
  infinity, $Z^{1} \sim \sin{\eta}\, e^{i \gamma}$ and $Z^{2} \sim
  \cos{\eta}\, e^{i \delta}$. In this way one manifestly recovers the wanted
  $\mathbb{R}\times S^{3}$ structure, where the $S^{3}$ is parametrized by
  $Z^{1}$ and $Z^{2}$ in terms of the coordinates $\eta,\delta$ and $\gamma$,
  while the $\mathbb{R}$ factor is parametrized by the phase $t$ of $Z^{0}$.
  The shift Eq.~(\ref{eq:tshift}) is necessary because, for $k=0$, the $t$
  coordinate in Eq.~(\ref{eq:firstconformalmetrick0}) is not the phase of
  $Z^{0}$, but was shifted to remove an additional term in $\chi$.  }

\begin{equation}
  \begin{array}{rcl}
\psi 
& = &
{\displaystyle
\frac{2 \cos{\eta}
\sin{(\delta-gt/\sqrt{3})}}{3-4\cos{(\delta-gt/\sqrt{3})}\cos{\eta}+\cos{2\eta}}\, ,
}
\\
& & \\
\zeta^{2} 
& = &
{\displaystyle
\frac{4 \sin^{2}{\eta}}{3-4\cos{(\delta-gt/\sqrt{3})}\cos{\eta}+\cos{2\eta}}\, ,
}
\\
& & \\
\varphi 
& = & 
{\displaystyle
\gamma- gt/\sqrt{3}
-\arccot{\left[ \cot(\delta-gt/\sqrt{3})-\frac{1}{\cos{\eta} \sin(\delta-gt/\sqrt{3})}
\right]}\, ,
}
\\
\end{array}
\end{equation}

\noindent
getting

\begin{equation}
\tilde{ds}{}^{2} 
= 
\frac{1}{3-4\cos(\delta-gt/\sqrt{3})\cos{\eta}+\cos{2\eta}}
\left[
d(gt/\sqrt{3})^{2}-d\eta^{2}-\cos^{2}{\eta}\, d\delta^{2}-\sin^{2}{\eta}\,
d\gamma^{2} 
\right]\,,
\end{equation}

\noindent
which again is conformal to $\mathbb{R}\times S^{3}$. The total conformal
factor is now 

\begin{equation}
\Omega 
= \cos{\xi}\, \sqrt{3-4\cos(\delta-gt/\sqrt{3})\cos{\eta}+\cos{2\eta}}\, ,
\end{equation}

\noindent
and, again, leads to a description of the boundary in two separate pieces. The
analysis if this case is much more involved and we will leave it for future work.

\section{Solutions}
\label{sec-solutions}

In this section we are going to try to solve
Eqs.~(\ref{eq:pure_maxwell_LI})-(\ref{eq:L}) to find supersymmetric solutions
of minimal gauged 5-dimensional supergravity.

We are going to search for solutions in which the functions $H,L,M,K$ only
depend on the coordinate $x^{2}$ which will play the r\^ole of ``radial''
coordinate and will be denoted by $\varrho$ as in the previous section. This
is possible if $W^{2}$ factorizes as follows

\begin{equation}
\label{eq:WversusPsiandPhi}
W^{2}=\Psi(\varrho)\Phi(x^{1},x^{3})\, ,
\end{equation}

\noindent
where, in order to solve Eq.~(\ref{eq:integrabilityagain}) $\Psi$ must take the
form

\begin{equation}
H=(\alpha\varrho+\beta)/\Psi\, ,
\end{equation}

\noindent
for some constants $\alpha$ and $\beta$. When $\alpha\neq 0$ we can eliminate
$\beta$ by shifting $\varrho$ and we can set $\alpha$ to $1$ by rescaling
$\varrho$. However, if $\alpha=0$, we cannot eliminate completely $\beta$: at
most we can set it to $1$ by rescaling $\Psi$. Thus, there are two possible
cases to be considered that we can parametrize with $\epsilon=0,1$:

\begin{equation}
\label{eq:HversusPsi}
H=\varrho^{\epsilon}/\Psi\, .
\end{equation}

If we assume that the metric function $\hat{f}$ is a function of $\varrho$
only, then it follows from Eq.~(\ref{eq:f-1}) that $\Phi$ is a solution of Liouville's equation

\begin{equation}
\left( \partial_{\underline{1}}^{2}+ \partial_{\underline{3}}^{2}\right)\log\Phi
=
-2 k\, \Phi\, ,
\end{equation}

\noindent
so that, for $k=1,0,-1$ it is given by the first of Eqs.~(\ref{eq:Phidef}),
then Eqs.~(\ref{eq:pure_integrabilityomegahat3})-(\ref{eq:L}) simplify
considerably: first,  Eq.~(\ref{eq:gK}) gives

\begin{equation}
\label{eq:gK2}
gK= \Psi^{\prime}/\Psi\, ,  
\end{equation}

\noindent
where primes denote derivation with respect to $\varrho$. Then,
Eq.~(\ref{eq:pure_integrabilityomegahat3}) $M$ can be integrated once to give

\begin{equation}
\label{eq:Mprime}
M^{\prime} = \frac{\alpha}{\Psi} -\sqrt{3}gL^{2}\, .
\end{equation}

\noindent
This result can be used to eliminate $M$ from Eq.~(\ref{eq:pure_maxwell_LI}),
giving

\begin{equation}
L^{\prime\prime} 
+\tfrac{4}{3}L^{\prime}\frac{\Psi^{\prime}}{\Psi}
+\tfrac{4}{3}L\frac{\Psi^{\prime\prime}}{\Psi} 
-\tfrac{2}{3}L\left(\frac{\Psi^{\prime}}{\Psi}\right)^{2}
-\tfrac{4}{\sqrt{3}}\alpha g\frac{\varrho^{\epsilon}}{\Psi^{2}}
=
0\, .
\end{equation}

\noindent
This equation has to be supplemented by Eq.~(\ref{eq:L}), that now takes the
form

\begin{equation}
\label{eq:LversusPhi}
L 
= 
\frac{\Psi}{8g^{2}\varrho^{\epsilon}} 
\left\{
-\frac{2k}{\Psi} 
-\tfrac{2}{3}\left(\frac{\Psi^{\prime}}{\Psi}\right)^{2}
+\frac{\Psi^{\prime\prime}}{\Psi}
\right\}\, .   
\end{equation}

Using the last equation to eliminate $L$ from the previous one, we get the promised
fourth order differential equation in $\Psi$

\begin{multline}
\label{eq:Psiequation}
-96\sqrt{3}\alpha g^{3} \varrho^{2+2\epsilon}
+4\varrho(\Psi^{\prime})^{2}\left( 3k\varrho-\epsilon\Psi^{\prime}
\right)
\\
+6\Psi\left[ -\epsilon(1+\epsilon)(\Psi^{\prime})^{2}
-4k\varrho^{2} \Psi^{\prime\prime}+2\epsilon\varrho\Psi^{\prime}
\left( 2k+\Psi^{\prime\prime} \right) \right]
\\
+9\Psi^{2}\left\{ \epsilon \left( 1+\epsilon \right)\Psi^{\prime\prime}
-2\epsilon\left[  k(1+\epsilon)+\varrho \Psi^{\prime\prime\prime} \right]+\varrho^{2}\Psi^{\prime\prime\prime\prime} \right\}
=
0\, .
\end{multline}

It is convenient to study the $\epsilon=0$ and $\epsilon=1$ cases
separately. The respective equations take the form

\begin{equation}
\label{eq:Psiequationepsilon0}
\epsilon=0\, ,
\,\,\,\,
\Rightarrow
\,\,\,\,
-32\sqrt{3}\alpha g^{3}+4
k(\Psi^{\prime})^{2}-8k\Psi\Psi^{\prime\prime}+3\Psi^{2}\Psi^{\prime\prime\prime\prime} 
=
0\, ,
\end{equation}

\noindent
and 

\begin{equation}
\label{eq:Psiequationepsilon1}
\begin{array}{rcl}
\epsilon=1\, ,
\,\,\,\,
\Rightarrow
\,\,\,\,
96\sqrt{3}\alpha g^{3} \varrho^4+4\varrho(\Psi^{\prime})^{2} 
\left( -3 k\varrho+\Psi^{\prime} \right)
+12\Psi\left( 2k\varrho-\Psi^{\prime} \right)
\left( -\Psi^{\prime}+\varrho\Psi^{\prime\prime} \right)
& & \\ & & \\
+9\Psi^{2}\left(4k-2\Psi^{\prime\prime}+2\varrho
\Psi^{\prime\prime\prime}-\varrho^{2}\Psi^{\prime\prime\prime\prime} \right)
& = &
0\, .
\end{array}
\end{equation}

Our experience with the RN-AdS$_{5}$ solutions in
Section~\ref{sec-RNADS5example} suggests the use of a polynomic Ansatz to
solve Eqs.~(\ref{eq:Psiequationepsilon0}) and (\ref{eq:Psiequationepsilon1}):

\begin{equation}
 \Psi=\sum_{n=0}^{N}c_{n} \varrho^{n}\, .
\end{equation}

In both equation, the term of highest order in $\varrho$ is always
proportional to the coefficient $c_{N}$ term in $\Psi$ and this term only
vanishes if $c_{N}=0$ or if $N\leq 3$, implying that $\Psi$ is at most of 3rd
order.

Let us analyze the $\epsilon=1$ and $\epsilon=0$ cases separately.

\subsection{The $\epsilon=1$ case: rotating black holes}
\label{sec-e=1}

Eq.~(\ref{eq:Psiequationepsilon1}) ($\epsilon=1$) is only solved if either
$c_{0}=c_{1}=0$ or if $c_{2}=k+\frac{c_{1}{}^{2}}{3 c_{0}}$. A parametrization
of the solution in terms of three parameters $a,b,c$ that covers both possibilities 
is\footnote{For $k=1$ this is the same as a solution found in \cite{Cassani:2015upa}
for a particular case of a scaling limit of the orthotoric base ansatz the authors use. The other
particular case of this limit analyzed in the same paper leads to a 
solution which was already known \cite{Figueras:2006xx} and that includes all known 
supersymmetric black hole solutions. As it turns out, this  very general solution
also includes our solution for $b=0$ and all three values of $k$, even if this was 
not shown explicitly in those papers. Our approach is 
in any case more systematic, since we have shown that these are the only possible 
solutions with polynomial $\Psi$ compatible with the assumptions we made.}
\begin{equation}
\Psi
=
\frac{1}{a}
\left[ c\varrho^{3}+\varrho^{2}+b\varrho+\frac{b^{2}}{3(1-a k)} \right]\, ,
\end{equation}

\noindent
and the constant $\alpha$ in Eq.~(\ref{eq:Mprime}) is constrained to take the
value

\begin{equation}
\alpha
=
\frac{1+3a k+3 b c\left[ \frac{3 b c}{1-a k}
-2\left( 1-2 a k\right)\right]}{24 \sqrt{3}a^{3} g^{3}}\, .
\end{equation}

Given the above values of $\Psi$ and $\alpha$, one can immediately compute
$W^{2}$ using Eqs.~(\ref{eq:WversusPsiandPhi}) and (\ref{eq:Phidef}), $H$
using Eq.~(\ref{eq:HversusPsi}) (with $\epsilon=1$), $L$ using
Eq.~(\ref{eq:LversusPhi}), $K$ using Eq.~(\ref{eq:gK2}) and $M$ using
Eq.~(\ref{eq:Mprime}). The latter, in particular, being the solution of a
first-order differential equation, contains an additional integration constant
that we call $d$.


The functions that appear in the metric and 1-form field are 

\begin{equation}
\label{eq:generalsolution}
\begin{array}{rcl}
\hat{f} 
& = & 
{\displaystyle
\frac{4a}{c}(g/\sqrt{3})^{2}
\frac{\varrho}{\varrho+\frac{(1-ak)}{3c}}\, ,
}
\\
& & \\
H 
& = & 
{\displaystyle
\frac{a\varrho}{c \varrho^{3}+\varrho^{2}+b \varrho+\frac{b^{2}}{3(1-ka)}}\, ,
}
\\
& & \\
W^{2}H
& = &
\varrho\, \Phi_{(k)}\, ,
\\
& & \\
\chi
& = &
\chi_{(k)}\, ,
\\
& & \\
\omega_{z}
& = &
{\displaystyle
d+\frac{b\left[9c\varrho+2(1-ak)\right]+3 \varrho\left[6 c^{2} \varrho^{2}
+3 c \varrho (2-ak)+(1-ak)^{2}\right]}{16\sqrt{3}g^{3} a^{2} \varrho^{2}}\, ,
}
\\
& & \\
\omega 
& = & 
{\displaystyle
-\frac{3\sqrt{3} c k}{16 g^{3} a}\chi_{(k)}\, .
}
\end{array}
\end{equation}

Notice that, since $\omega$ is given by a constant times $\chi_{(k)}$, it can be
reabsorbed in $\omega_{z}$ with a shift in the $t$ coordinate, so that 

\begin{equation}
\label{eq:omegaz}
\omega = 0\, ,
\hspace{1cm}
\omega_{z}
=
d
+\frac{18c^{2}\varrho^{3}+18c(1-ak)\varrho^{2} 
+[9bc+3(1-ak)^{2}]\varrho +2b(1-ak)}{16\sqrt{3}g^{3} a^{2}
\varrho^{2}}\, .
\end{equation}

Notice also that the full 5-dimensional metric is invariant under the
rescaling $t\rightarrow t/\alpha$, $\varrho\rightarrow \alpha \varrho$,
$b\rightarrow \alpha b$, $c\rightarrow c/\alpha$, $d\rightarrow
d/\alpha$. This allows to set one of the constants $b,c,d$ to $1$, provided it
is not zero, leaving only three independent parameters. Then, assuming $c\neq 0$ 
(the $c=0$ case will be dealt with later) we can use this freedom to 
normalize the metric so that $\hat{f}\rightarrow 1$ for large values of
$\varrho$, setting

\begin{equation}
\frac{4a}{c}(g/\sqrt{3})^{2}=1\, .  
\end{equation}

Eliminating in this way $c$ from the non-vanishing functions that define the family of
solutions, we get

\begin{equation}
\begin{array}{rcl}
\hat{f} 
& = & 
{\displaystyle
\varrho
\left[\varrho+\frac{(1-ak)}{4ag^{2}}\right]^{-1}
}
\\
& & \\
H 
& = & 
{\displaystyle
\varrho
\left\{
\frac{4g^{2}}{3} \varrho^{3}+\frac{1}{a}\varrho^{2}
+\frac{b}{a} \varrho+\frac{b^{2}}{3a(1-ka)}
\right\}^{-1}\, ,
}
\\
& & \\
W^{2}H
& = &
\varrho\, \Phi_{(k)}\, ,
\\
& & \\
\chi
& = &
\chi_{(k)}\, ,
\\
& & \\
\omega_{z}
& = &
{\displaystyle
d+
\varrho^{-2}
\left\{
\frac{2g}{\sqrt{3}}\varrho^{3} +\frac{\sqrt{3}(1-ak)}{2ag}\varrho^{2}
+\left[
\frac{\sqrt{3}b}{4ag}
+\frac{\sqrt{3}(1-ak)^{2}}{16a^{2}g^{3}}
\right]\varrho 
+\frac{b(1-ak)}{8\sqrt{3}a^{2}g^{3}}
\right\}
}
\, .
\end{array}
\end{equation}

Comparing this family of solutions with the supersymmetric RN-AdS$_{5}$
solution in Eqs.~(\ref{eq:RNAdS5k=1}) we find that the latter are included in
the former for the following values of the independent integration constants:

\begin{equation}
 k=1\, ,
\hspace{1cm} 
a=(1+\sqrt{3}{g}^{2}q)^{-1}\, ,
\hspace{1cm} 
b=\frac{3g^{2}q^{2}}{4(1+\sqrt{3}{g}^{2}q)}\, ,
\hspace{1cm}  
d=0\, .
\end{equation}

Since the integration constant $d$ is independent of the rest, we could extend
the RN-AdS$_{5}$ solution by switching it on. The resulting solutions are no
longer asymptotically AdS$_{5}$. This is true for the whole family of
solutions presented here and, henceforth, we will set $d=0$ in what follows.

Taking

\begin{equation}
a^{-1}\rightarrow k\, ,
\hspace{1cm} 
b=d=0\, ,
\end{equation}

\noindent
we get the 3 different forms of the AdS$_{5}$ metrics Eq.~(\ref{eq:ads5k}).

Perhaps more interestingly, for 

\begin{equation}
k=1\, ,
\hspace{1cm} 
a=\frac{1}{4 \alpha^{2}}\, ,
\hspace{1cm} 
b=d=0\, ,
\end{equation}

\noindent
one recovers the asymptotically-AdS$_{5}$, supersymmetric, charged, rotating
black holes found in Ref.~\cite{Gutowski:2004ez}.\footnote{To compare our
  solution with the solution described in Section~4.1 of
  Ref.~\cite{Gutowski:2004ez} we first have to identify the constants
  $g/\sqrt{3}=\ell^{-1}$, transform our radial coordinate $x^{2}\equiv\varrho
  = \alpha^{2}\ell^{2}\sinh{(\rho/\ell)}$ and our isometric coordinate
  $z=\phi$. Furthermore, we have to make the usual coordinate transformation
  to go from conformally flat coordinates $x^{1}, x^{3}$ to spherical ones
  ($\theta,\psi$ in Ref.~\cite{Gutowski:2004ez}) on the 2-sphere:
  $x^{1}+ix^{3}=\tan{(\theta/2)}\, e^{i\psi}$.} The mass $M$, the
non-vanishing angular momentum $J$ and the electric charge $q$ of these
solutions are given in terms of the only independent parameter $a$
by\footnote{Our normalization of the electric charge differs by a factor of
  $2$ from that of Ref.~\cite{Gutowski:2004ez}, so, for $J=0$, we get
  Eq.~(\ref{eq:susyrel}). Furthermore, we remind the reader that we have
  chosen units such that Eq.~(\ref{eq:units2}) holds.}

\begin{equation}
M
= 
R_{0}^{2} +\frac{g^{2}}{2}R_{0}^{4} + \frac{2g^{4}}{27}R_{0}^{6}\, ,
\hspace{1cm}
J
= 
\frac{g}{2\sqrt{3}}R_{0}^{4} +\frac{g^{3}}{9\sqrt{3}}R_{0}^{6}\, ,
\hspace{1cm}
q
=
\tfrac{1}{\sqrt{3}}R_{0}^{2} +\frac{g}{6\sqrt{3}}R_{0}^{4}\, ,
\end{equation}

\noindent
where $R_{0}^{2}= (1-a)/(ag^{2})$,  so that

\begin{equation}
\label{eq:BPS_bound}
M-\frac{2g}{\sqrt{3}}|J| = \sqrt{3}|q|\, .  
\end{equation}

These black-hole solutions have a regular near-horizon geometry, and the
horizon is a squashed 3-sphere. This is just one of the three possible
near-horizon geometries found in Ref.~\cite{Gutowski:2004ez}. We are going to
show that there are $k=0,-1$ solutions which have the other two near-horizon
geometries. In particular, for 
\begin{equation}
 k=0\, ,
\hspace{1cm}
b=d=0\, ,
\end{equation}
the remaining parameter $a$ can be set to $1$ with a rescaling of the 
coordinates and one gets the solution obtained in Ref.~\cite{Gutowski:2004ez}
as the ``large black-hole limit'' ($R_{0}\rightarrow \infty$) of the $k=1$
solution.

\subsubsection{Near-horizon geometries}
\label{sec-NHgeometries}

The event horizon, if it exists, must be placed at $\varrho=0$. When 
the parameter $b\neq 0$, our experience with the supersymmetric
RN-AdS$_{5}$ solution suggests that there is no event horizon and the
$\varrho\rightarrow 0$ limit is not a near-horizon geometry even if it is a
regular one. Therefore, we are going to study separately the $b=0$ and $b\neq
0$ cases.

When $b=0$, defining first the coordinates $u,v$

\begin{equation}
\label{eq:NHchangeofcoordinates}
\begin{array}{rcl}
dt 
& = &
{\displaystyle
\mp \frac{(1-ak)^{1/2}(1+3ak)^{1/2}}{4ag}
\left[
du
-
\frac{(1-ak)}{4g^{2}}\frac{d\varrho}{\varrho^{2}}  
\right]\, ,
}
\\
& & \\
dz
& = &
{\displaystyle
dv \pm \sqrt{3}a\frac{(1-ak)^{1/2}}{(1+3ak)^{1/2}}\frac{d\varrho}{\varrho}\, ,
}
\\
\end{array}
\end{equation}

\noindent
the near-horizon geometry can be written in a form that generalizes
the one obtained for the $k=1$ case in Ref.~\cite{Gutowski:2004ez}

\begin{multline}
\label{eq:generalNHmetric}
ds^{2}
=
\Delta^{2} \varrho^{2}du^{2} -2dud\varrho
+\frac{6k\Delta}{\ell(\Delta^{2}-3\ell^{-2})}\varrho du (dv+\chi_{(k)})
\\ \\
- \frac{k}{\Delta^{2}-3\ell^{-2}}
\left[
\frac{k\Delta^{2}}{\Delta^{2}-3\ell^{-2}}(dv+\chi_{(k)})^{2} 
+d\Omega^{2}_{(2,k)}
\right]\, ,
\end{multline}

\noindent
where $3\ell^{-2}=g^{2}$ and we have defined

\begin{equation}
\label{eq:Delta}
\Delta^{2}  = \frac{1+3ak}{1-ak} g^{2}\, .
\end{equation}

\noindent
Observe that the combination $k/(\Delta^{2}-3\ell^{-2})= (1-ak)/(4ag^{2})$
does not vanish for $k=0$. it does not become negative for $k=-1$ either. 

For the $k=0$ case, a rescaling of the coordinates $v\equiv 4g
\omega$, $x^{1}\equiv g x$, $x^{3}\equiv gy$ brings the
above near-horizon metric into the form

\begin{multline}
ds^{2}
=
\frac{3}{\ell^{2}} \varrho^{2}du^{2} -2dud\varrho
+\frac{6}{\ell}\varrho du \left[dv^{\prime}+\frac{\sqrt{3}}{2\ell}(ydx-xdy)\right]
\\ \\
-\left[dv^{\prime}+\frac{\sqrt{3}}{2\ell}(ydx-xdy)\right]^{2}
-dx^{2} -dy^{2}\, ,
\end{multline}

\noindent
which, at $\varrho=0$ gives the standard metric of the homogeneous
\textit{Nil} group manifold and which, upon dimensional reduction along
$v^{\prime}$ gives the metric of AdS$_{2}\times\mathbb{E}^{2}$.\footnote{The
  solution in Eqs.~(4.62) of Ref.~\cite{Gutowski:2004ez}, which corresponds 
  to our $k=b=d=0$ solution, obtained as the
  ``large black-hole limit'' ($R_{0}\rightarrow \infty$) of the $k=1$
  solution, has a horizon with precisely this near-horizon geometry. However,
  the solution was considered by the authors to be not
  asymptotically-AdS$_{5}$ because, the asymptotic geometry, written in
  Eq.~(4.63) of that reference, was interpreted as a supersymmetric
  plane-fronted wave. It is not difficult to see that, actually, is the $k=0$
  form of AdS$_{5}$ as given in Eq.~(\ref{eq:ads5k}) upon the coordinate
  change $\varrho=S^{2}$. Thus, the ``large black-hole limit'' of the $k=+1$
  black hole gives the $k=0$ black hole.}

For $k=-1$ we rescale $v\equiv -(\Delta^{2}-3\ell^{-2}) v^{\prime}/\Delta$ to
obtain

\begin{multline}
ds^{2}
=
\Delta^{2} \varrho^{2}du^{2} -2dud\varrho
+\frac{6}{\ell}\varrho du 
\left[dv^{\prime}-\frac{\Delta}{\Delta^{2}-3\ell^{-2}}\chi_{(k)}\right]
\\ \\
- \left[dv^{\prime}-\frac{\Delta}{\Delta^{2}-3\ell^{-2}}\chi_{(k)}\right]^{2} 
+\frac{\Delta}{\Delta^{2}-3\ell^{-2}}d\Omega^{2}_{(2,-1)}\, ,
\end{multline}

\noindent
which, upon dimensional reduction along $v^{\prime}$ gives the metric of
AdS$_{2}\times\mathbb{H}_{2}$ and which, for $\Delta=0$, gives the metric of
AdS$_{3}\times\mathbb{H}_{2}$ which arises as the near-horizon geometry of the
black strings of Ref.~\cite{Klemm:2000nj}.

When $b\neq 0$ we obtain in the $\varrho\rightarrow 0$ limit a completely
regular geometry that does not correspond to a horizon:

\begin{multline}
ds^{2}
=
\frac{16 a^{2} g^{4} \varrho^{2}}{(1-ak)^{2}}dt^{2}
+\frac{4 b g}{\sqrt{3}(1-ak)}dt(dz+\chi_{(k)}) 
-\frac{3(1-ak)^{2}}{4 b^{2}g^{2}}d\varrho^{2}\\ \\
\\ 
-\frac{1-ak}{4 a g^{2}}
\left\{
\frac{9 b^{2} c^{2}+(1-ak)^{3}(1+3ak)-6bc(1-ak)^{2}}{4 a(1-ak)^{3}}  (dz+\chi_{(k)})^{2}
+d\Omega_{(2,k)}^{2}
\right\}\, .
\end{multline}

\subsubsection{Asymptotic limits}

The naive asymptotic limits of these solutions are the different forms of
AdS$_{5}$ presented in Section~\ref{sec-ADS5example}. However, these limits
are very subtle and must be analyzed using the same methods we used for the
different forms of AdS$_{5}$ at the end of Section~\ref{sec-ADS5example}. We
can use exactly the same changes of coordinates. Then, it can be seen that
only for $c\neq0$ and $d=0$ these solutions can be asymptotically AdS$_{5}$.

Notice however that for $k=0,-1$ the conformal 4-dimensional metric presents a
singularity where the conformal factor multiplying the
$\mathbb{R}\times$S$^{3}$ metric diverges. For pure AdS$_{5}$ this problem can
of course be solved by simply changing to the usual ($k=1$) global
coordinates, which amounts to taking a slightly different asymptotic limit.
For the full solutions however this is not the case: we have verified that for
$k=-1$ a simple $k=1$ AdS$_{5}$-like coordinate transformation leads to a Weyl
tensor that diverges in $\eta=\frac{\pi}{2}$ for all values of $\xi$, while it
vanishes in $\xi=\frac{\pi}{2}$ if $\eta\neq\frac{\pi}{2}$.  The situation can
be improved with a modified coordinate transformation giving a regular Weyl
tensor, which however still diverges as one approaches
$(\xi,\eta)=(\frac{\pi}{2},\frac{\pi}{2})$. This could indicate that these
solutions do not asymptote to AdS$_{5}$ globally, but only locally.

The $k=0$ case is more complicated, due to the more involved transformation
between the correspondent AdS parametrization and the $k=1$ one, and we have 
not studied this case in detail. One could expect however a similar behaviour as
in the $k=-1$ case.

\subsubsection{The conserved charges}

For $k=1$ we can compute the conserved charges of the solutions following the
prescription given in \cite{Ashtekar:1999jx}. The mass is given by the
conserved charge associated to the Killing vector pointing along the time
direction of the conformal boundary $\mathbb{R}\times$S$^{3}$ metric, and is

\begin{equation}
M
=
\frac{-31 a^{4} + a^{3} (43-76 g^{2} b) 
+a^{2} (3+44 g^{2} b-64 g^{4} b^{2})+a(-11+32 g^{2} b)-4}{54 g^{2} a^{3} (a-1)}\, .
\end{equation}

\noindent
The angular momenta associated to $\partial_{\psi}$ and $\partial_{\phi}$ are

\begin{equation}
J_{\psi}
=
\frac{\left[a^{2}-2a(1+2g^{2}b)+1\right]\left[ 7 
a^{2}+a(-5+8 g^{2} b)-2 \right]}{18\sqrt{3}a^{3} g^{3} (a-1)}\, ,
\qquad 
J_{\phi}=0\,.
\end{equation}

The electric charge can be computed integrating the Hodge dual of the 
gauge field strength over a 3-sphere at infinity, and is given by

\begin{equation}
Q
=
\frac{-5 a^{2}+4a(1-g^{2}b)+1}{6\sqrt{3}a^{2} g^{2}}\,.
\end{equation}

It is straightforward to verify that these expressions reduce to the expected
values for the Gutowski-Reall black hole and for RN-AdS$_{5}$, and that the
BPS bound Eq.~(\ref{eq:BPS_bound}) is satisfied for all values of the
parameters $a$ and $b$.

\subsubsection{A homogeneous solution?}

Besides the three different parametrizations of AdS$_{5}$ mentioned above,
there is another choice of the parameters giving a solution apparently
free of curvature singularities, for which in particular the Ricci and 
Kretschmann scalars and the Ricci tensor fully contracted with itself 
are all constant. It is given by $k=0$, $d=0$ and $b=\frac{1}{3c}=\frac{1}{4 a g^{2}}$,
and its metric, after a rescaling of the coordinates, is given by

\begin{equation}
 ds^{2}=\frac{3}{4g^{2}}\left[\frac{\varrho^{2}}{(1+\varrho)^{2}}dt^{2}+2(1+\varrho)dt(dz+\chi_{(0)})-\frac{d\varrho^{2}}{(1+\varrho)^{2}}-(1+\varrho)d\Omega_{(0)}^{2}  \right]
\end{equation}

\noindent
with gauge field strength

\begin{equation}
 F=-\frac{3}{2g}\frac{d\varrho\wedge dt}{(1+\varrho)^{2}}\,.
\end{equation}

Since $b\neq0$ the solution is horizonless, and since $d=0$ it is 
asymptotically, at least locally, AdS$_{5}$.

In terms of a Vielbein

\begin{equation}
 F=-2g \frac{e^{2}\wedge (e^{0}-e^{1})}{\varrho}\,.
\end{equation}

\subsubsection{The $c=0$ solutions}

The $c=0$ solutions (with $d=b=0$) can be seen to coincide identically with
the near-horizon geometries recovered in Section~\ref{sec-NHgeometries}:
setting $b=c=d=0$ in Eqs.~(\ref{eq:generalsolution}) and (\ref{eq:omegaz}) we
get a metric that coincides exactly with that determined by the leading terms
in the $\varrho\rightarrow 0$ limit. The change of coordinates
Eq.~(\ref{eq:NHchangeofcoordinates}) and the replacement of the parameter $a$
by $\Delta$ defined in Eq.~(\ref{eq:Delta}) brings it into the form
Eq.~(\ref{eq:generalNHmetric}).  The near-horizon configuration is a
supersymmetric solution in its own right ad it is included in the general
solution that we have presented.

\subsection{The $\epsilon=0$ case: G\"odel universes}
\label{sec-e=0}

First of all, in this case, Eq.~(\ref{eq:constraintijcurved}) implies
$d\chi=0$, and one can set $\chi=0$. Thus, we can absorb any constant term in
$\omega_{z}$ in a redefinition on $t$. Furthermore, the integration constant
$\alpha$ in Eq.~(\ref{eq:Mprime}) must take the value

\begin{equation}
 \alpha=k\frac{c_{1}{}^{2}-4 c_{0} c_{2}}{8\sqrt{3} g^{3}}\, ,
\end{equation}

\noindent
and one has to distinguish between the $k=0$ case, in which $\Psi$ can be an
arbitrary 3rd order polynomial, and the $k\neq 0$ one, in which $c_{3}$ must
vanish, meaning that $\Psi$ must be of just 2nd order.

For $\epsilon=0$ and $k=0$ this gives (after the integration to obtain $M$)

\begin{equation}
\begin{array}{rcl}
\hat{f}^{-1} 
& = &
{\displaystyle
\frac{c_{2}+3c_{3}\varrho }{4 g^{2}}\, ,
}
\\
& & \\
H^{-1}
& = &
c_{0}+c_{1}\varrho+c_{2}\varrho^{2}+c_{3}\varrho^{3}\, ,
\\
& & \\
W^{2}H
& = &
\Phi_{(0)}\, ,
\\
& & \\
\chi
& = &
0\, ,
\\
& & \\
\omega_{z}
& = &
{\displaystyle
\frac{\sqrt{3}}{16 g^{3}}
\left[(c_{2}{}^{2}+3c_{1}c_{3})\varrho +6c_{2}c_{3}\varrho^{2} 
+6c^{2}_{3}\varrho^{3}\right]\, ,
}
\\
& & \\
\omega
& = &
{\displaystyle
-\sqrt{3}\frac{c_{2}{}^{2}-3 c_{1} c_{3}}{16 g^{3}} \chi_{(0)}\, ,
}
\\
\end{array}
\end{equation}

\noindent
so that, in particular, the metric takes the form

\begin{align}
ds^{2} 
& 
= 
\frac{16g^{4}}{(c_{2}+3c_{3}\varrho)^{2}}
\left\{
dt 
+\frac{\sqrt{3}}{16g^{3}}
\left[(c_{2}{}^{2}+3c_{1}c_{3})\varrho +6c_{2}c_{3}\varrho^{2} 
+6c^{2}_{3}\varrho^{3}\right] dz
\right.
\nonumber \\
&
\left.
+\frac{\sqrt{3}}{8g^{3}}(c_{2}^{2}-3c_{1}c_{3})(xdy-ydx)
\right\}^{2}
-\frac{1}{4g^{2}}(c_{2}+3c_{3}\varrho)
(c_{0}+c_{1}\varrho+c_{2}\varrho^{2}+c_{3}\varrho^{3})dz^{2}
\nonumber \\
&
-\frac{1}{4g^{2}}\frac{(c_{2}+3c_{3}\varrho)}{(c_{0}+c_{1}\varrho+c_{2}\varrho^{2}+c_{3}\varrho^{3})}
d\varrho^{2}
-\frac{1}{g^{2}}(c_{2}+3c_{3}\varrho)(dx^{2}+dy^{2})\, .
\end{align}

For $\epsilon=0$ and $k\neq0$, the functions that define the solution are
given by 

\begin{equation}
\begin{array}{rcl}
\hat{f}^{-1}
& = &
{\displaystyle
\frac{c_{2}-k}{4 g^{2}}\, ,
}
\\
& & \\
H^{-1}
& = &
c_{0}+c_{1}\varrho+c_{2}\varrho^{2}\, ,
\\
& & \\
W^{2}H
& = &
\Phi_{(k)}\, ,
\\
& & \\
\chi
& = &
0\, ,
\\
& & \\
\omega_{z}
& = &
{\displaystyle
-
\frac{\left( 3k-c_{2} \right)\left( k+3 c_{2} \right)}{16
  \sqrt{3}g^{3}}\varrho\, ,
}
\\
& & \\
\omega
& = &
{\displaystyle
\frac{\left( k-3 c_{2} \right)\left( 3k+c_{2}
  \right)}{16\sqrt{3}g^{3}}\chi_{(k)}\, ,
}
\end{array}
\end{equation}

\noindent
so that the metric, in particular, takes the form

\begin{align}
ds^{2}
& 
=
\frac{16g^{4}}{(c_{2}-k)^{2}}
\left\{
dt
-\frac{(3k-c_{2})(k+3c_{2})}{16\sqrt{3}g^{3}}\varrho dz  
+\frac{(3k+c_{2})(k-3c_{2})}{16\sqrt{3}g^{3}}\chi_{(k)}
\right\}^{2}
\nonumber \\
&
-\frac{(c_{2}-k)}{4g^{2}}(c_{0}+c_{1}\varrho+c_{2}\varrho^{2})dz^{2}
-\frac{(c_{2}-k)}{4g^{2}(c_{0}+c_{1}\varrho+c_{2}\varrho^{2})}d\varrho^{2}
-\frac{(c_{2}-k)}{4g^{2}}d\Omega^{2}_{(k)}\, .
\end{align}

The parameters of the solutions above can be reduced by shifting and 
rescaling $\varrho$. The remaining independent possibilities are:

\begin{enumerate}
 \item $k=0$, $c_{3}=1$, $c_{2}=0$, $c_{1}$ and $c_{0}$ arbitrary.
 \item $k=0,\pm1$, $c_{3}=0$, $c_{2}\neq 0$, $c_{2}>k$, $c_{1}=0$ and $c_{0}=0$.
 \item $k=0,\pm1$, $c_{3}=0$, $c_{2}\neq 0$, $c_{2}>k$, $c_{1}=0$ and $c_{0}=1$.
 \item $k=-1$, $c_{3}=0$, $c_{2}=0$, $c_{1}=1$ and $c_{0}=0$.
 \item $k=-1$, $c_{3}=0$, $c_{2}=0$, $c_{1}=0$ and $c_{0}=1$.
\end{enumerate}

Furthermore, for the cases 2. and 3., if $k=0$, it is possible to set
$c_{2}=1$.

\textbf{Case 1.} is in general of difficult interpretation, however if
$c_{1}=c_{0}=0$ the solution after a rescaling of the $t$ coordinate takes the
form

\begin{equation}
ds^{2}
=
\frac{3}{4g^{2}}\left[ \frac{dt^{2}}{\varrho^{2}}+2\varrho dt dz
-\frac{d\varrho^{2}}{\varrho^{2}}-\varrho d\Omega_{(0)}^{2} \right]
\end{equation}

\begin{equation}
 F=\frac{3}{2g}\frac{d\varrho}{\varrho^{2}}\wedge dt\, .
\end{equation}

The Ricci and Kretschmann scalars and the Ricci tensor fully contracted with
itself are constant for this metric, suggesting it may represent a homogeneous
space.  The gauge field strength is constant if expressed in terms of a
Vielbein, and represents a homogeneous electric field directed along
$\varrho$.

In all the remaining cases the abovementioned curvature scalars are
constant

\textbf{Cases 2.} and \textbf{3.}:

\begin{multline}
ds^{2}
=
\left( \frac{4g^{2}}{c_{2}-k} \right)^{2}
\left[ 
dt-\frac{(3k-c_{2})(k+3c_{2})}{16\sqrt{3}g^{3}c_{2}}\chi_{(-1)}
+\frac{(k-3c_{2})(3k+c_{2})}{16\sqrt{3}g^{3}}\chi_{(k)} 
\right]^{2}\\
\label{eq:gen_godel_metric}
-\frac{c_{2}-k}{4g^{2}c_{2}}\left[d\Omega_{(-1)}^{2}+c_{2}  d\Omega_{(k)}^{2} \right]\,.
\end{multline}

\textbf{Cases 4.} and \textbf{5.}:

\begin{equation}
ds^{2}
=
16 g^{4}
\left[
dt+\frac{\sqrt{3}}{16 g^{3}}\left( \chi_{(-1)}-\chi_{(0)}\right)  
\right]^{2}
-\frac{1}{4 g^{2}}\left[ d\Omega_{(0)}^{2}+ d\Omega_{(-1)}^{2} \right]\, .
\end{equation}

The general expression of the gauge field strength for $c_{3}=0$ is

\begin{equation}
F
=
\frac{1}{4g(c_{2}-k)}\left[ \left( 3k^{2}+4 c_{2} k+ c_{2}^{2}\right)d\varrho\wedge dz 
+\left( k^{2}+4kc_{2}+3 c_{2}^{2} \right)\Phi_{(k)}dx^{3}\wedge dx^{1}\right]\,.
\end{equation}
Notice that the metric and gauge field for cases 4. and 5. can actually 
be seen as the particular case $k=0$ of the ones for cases 2. and 3., so that all
cases with $c_3=0$ have metric (\ref{eq:gen_godel_metric}) and gauge field strength
that can be rewritten as
\begin{equation}
 F
 =
 \frac{1}{4g c_2 (c_{2}-k)}\left[ \left( 3k^{2}+4 c_{2} k+ c_{2}^{2}\right)d\chi_{(-1)} 
+\left( k^{2}+4kc_{2}+3 c_{2}^{2} \right)c_2 d\chi_{(k)}\right]\,.
\end{equation}

These solutions are 5-dimensional supersymmetric generalizations of the
4-dimensional G\"odel's rotating universe \cite{Godel:1949ga}, which also
solves Einstein's equations with a cosmological constant and contain the
2-dimensional metric $d\Omega^{2}_{(-1)}$ and the associated 1-form
$\chi_{(-1)}$. As in that case and also in the case of the 5-dimensional
G\"odel solution of the ungauged theory
\cite{Tseytlin:1996as,Gauntlett:2002nw}, the solution contains closed timelike
curves. Those solutions are also homogeneous spaces and it would be
interesting to know if the three solutions presented share this property, as
the constancy of their curvature invariants seems to indicate. In the
ungauged 5-dimensional case \cite{Meessen:2004mh}, the dimensional reduction
over the time direction gives rise to a solution of Euclidean
$\mathcal{N}=2,d=4$ supergravity with an anti-selfdual Abelian instanton field
and a geometry which, instead of $\mathbb{E}^{4}$ is given by
$\mathbb{H}_{2}\times (S^{2},\mathbb{E}^{2},\mathbb{H}_{2})$ geometry. It is
also likely that these 3 G\"odel solutions can be obtained from the 3
near-horizon geometries discussed above by the limiting procedure proposed in
Ref.~\cite{Meessen:2004mh}, since the standard Penrose limit cannot be used in
gauged supergravity.\footnote{We thank P.~Meessen for comments on this point.}

\section{Reduction to $d=4$}
\label{sec-reduction}

The dimensional reduction over a circle of the theory of minimal 5-dimensional
supergravity gives a theory of $\mathcal{N}=2,d=4$ supergravity coupled to one
vector multiplet and determined by the cubic prepotential
$\mathcal{F}=-(\mathcal{X}^{1})^{3}/\mathcal{X}^{0}$. The complex scalar
$t\equiv -\mathcal{X}^{1}/\mathcal{X}^{0}$ parametrizes an
SL$(2,\mathbb{R})/$SO$(2)$ $\sigma$-model with K\"ahler potential
$e^{\mathcal{K}}
=(\Im\mathfrak{m}\, t)^{3}$. The relation between this
and the rest of the 4-dimensional fields and the 5-dimensional ones (for which
we use hats here: $\hat{g}_{\hat{\mu}\hat{\nu}}$ and $\hat{A}_{\hat{\mu}}$,
  where $\hat{\mu}=\mu,z$) is given by

\begin{eqnarray}
g_{\mu\nu}
& = &
k\left(\hat{g}_{\mu\nu}
+\frac{\hat{g}_{\mu\underline{z}}\hat{g}_{\nu\underline{z}}}{k^{2}} \right)\, ,
\\
& & \nonumber \\
A^{0}{}_{\mu}
& = &
-\tfrac{1}{2\sqrt{2}}\frac{\hat{g}_{\mu\underline{z}}}{k^{2}}\, ,
\\
& & \nonumber \\
A^{1}{}_{\mu}
& = &
-\tfrac{1}{2\sqrt{6}}\hat{A}_{\mu}
+\tfrac{1}{\sqrt{3}}\hat{A}_{\underline{z}}A^{0}{}_{\mu}\, ,
\\
& & \nonumber \\
t 
& = &  
\tfrac{1}{2\sqrt{3}} \hat{A}_{\underline{z}} +\tfrac{i}{2} k\, ,
\end{eqnarray}

\noindent
where

\begin{equation}
k^{2} = -\hat{g}_{\underline{z}\underline{z}}\, ,
\end{equation}

\noindent
is the Kaluza-Klein (KK) scalar measuring the local size of the compactification
circle. It is assumed to be positive so the isometric coordinate $z$ is
spacelike.

The dimensional reduction of bosonic sector of the minimal, gauged,
5-dimensional theory of supergravity gives exactly the same action with the
same relations between the 5- and the 4-dimensional fields except for an
additional term corresponding to the 5-dimensional constant.\footnote{In
  presence of hypermultiplets one can get additional terms in the scalar
  potential using generalized dimensional reduction \cite{Gaddam:2014mna}.}
In $d=4$ it appears multiplied by the KK scalar and becomes a
negative-definite (but unbound) scalar potential. Taking into account the
relation between the 5- and the 4-dimensional gauge coupling constants
$g=-g_{4}/\sqrt{24}$, the 4-dimensional scalar potential is

\begin{equation}
V_{4} = -(g_{4}/\sqrt{3})^{2}(\Im\mathfrak{m}\, t)^{-1}\, .
\end{equation}

This potential does not have any extremum at regular points of the scalar
manifold and, therefore, the theory does not admit an AdS$_{4}$
vacuum.\footnote{In the dimensional reduction of a non-minimal gauged theory
  with vector supermultiplets and a 5-dimensional scalar potential
  $V_{5}(\phi)$ we obtain a 4-dimensional scalar potential which will always
  be of the form
\begin{equation}
V_{4} = k^{-1}V_{5}(\phi)\, ,  
\end{equation}
and analogous observations apply as well.  }  The most symmetric
supersymmetric vacuum solution is probably the one obtained by dimensional
reduction of the AdS$_{5}$ which we are going to review shortly. Since
AdS$_{5}$ is the only maximally supersymmetric solution of minimal, gauged,
5-dimensional supergravity, this is only solution that could be maximally
supersymmetric in the 4-dimensional theory.\footnote{Any maximally
  supersymmetric solution of the 4-dimensional solution must necessarily
  correspond to a maximally supersymmetric solution of the 5-dimensional
  theory. The converse is not true.} All the asymptotically-AdS$_{5}$
solutions become 4-dimensional solutions that have the same asymptotic
behaviour as that solution.

Using the above rules for the dimensional reduction, the metric and 2-form
potential of the timelike supersymmetric solutions give rise to the following
4-dimensional fields:

\begin{eqnarray}
ds^{2}
& = &
e^{2U}(dt+\omega)^{2}
-e^{-2U}\gamma_{\underline{r}\underline{s}}dx^{r}dx^{s}\, ,
\\
& & \nonumber \\
A^{0}
& = &
\tfrac{1}{2\sqrt{2}} 
\left\{
-\frac{\hat{f}^{2}\omega_{z}}{k^{2}} (dt+\omega) +\chi
\right\}\, ,
\\
& & \nonumber \\
A^{1}
& = &
-\tfrac{1}{2\sqrt{6}} 
\left\{
\frac{\hat{f}^{2}\omega_{z}}{k^{2}} 
\left[
-\sqrt{3}\hat{f}\omega_{z} +\frac{\partial_{\underline{2}}\log{W^{2}}}{2gH}
\right]
(dt+\omega) 
\right\}\, ,
\nonumber \\
& & \nonumber \\
& & 
-\frac{1}{2g}\left(\partial_{\underline{1}}\log{W^{2}}dx^{3} 
-\partial_{\underline{3}}\log{W^{2}}dx^{1} \right)\, ,
\\
& & \nonumber \\
t
& = &
\tfrac{1}{2}
\left[
-\hat{f}\omega_{z} +\frac{\partial_{\underline{2}}\log{W^{2}}}{2\sqrt{3}gH}
\right]
+\tfrac{i}{2}k\, ,
\end{eqnarray}

\noindent
where

\begin{eqnarray}
k^{2}
& = &
\hat{f}^{-1}H^{-1}-\hat{f}^{2}\omega_{z}^{2}\, ,
\\
& & \nonumber \\
\gamma_{\underline{r}\underline{s}}dx^{r}dx^{s}
& = &
(dx^{2})^{2} +W^{2}[(dx^{1})^{2}+(dx^{3})^{2}]\, ,  
\\
& & \nonumber \\
e^{-2U}
& = &
k\hat{f}^{-1}H
=
\sqrt{HL^{3} +\tfrac{1}{16}L^{2}K^{2} -M^{2}H^{2} -\tfrac{\sqrt{3}}{2}MLKH
  +\tfrac{1}{12\sqrt{3}}MK^{3}}\, ,
\end{eqnarray}

\noindent
and $H,K,L,M,W,\omega$ and $\chi$ are the same functions and 1-form that occur
in the 5-dimensional metric. The functions $H,K,L,M$ can be identified with
the building blocks of the 4-dimensional timelike supersymmetric solutions
(harmonic functions on $\mathbb{E}^{3}$ in the ungauged case).

The most interesting examples we can apply these relations to are AdS$_{5}$
and the Gutowski-Reall black hole.\footnote{The dimensional reduction of the
  supersymmetric Reissner-Nordstr\"om-AdS$_{5}$ solution gives a singular
  solution.}

\subsection{Reduction of AdS$_{5}$}
\label{sec-reductionofAdS5}

Applying the above relations to the $k=1$ supersymmetric form of AdS$_{5}$ in
Eq.~(\ref{eq:ads5k1}) we get the 4-dimensional solution 

\begin{eqnarray}
ds^{2}
& = & 
\varrho^{1/2}(1+\tfrac{1}{18}g_{4}^{2}\varrho)dt^{2} 
-\frac{d\varrho^{2}}{\varrho^{1/2}(1+\tfrac{1}{18}g_{4}^{2}\varrho)}  
-\varrho^{3/2}d\Omega^{2}_{(2,1)}\, ,
\\
& & \nonumber \\
A^{0}
& = &
\frac{1}{2\sqrt{2}}\chi_{(1)}\, ,
\\
& & \nonumber \\
A^{1}
& = &
\frac{1}{g_{4}}\chi_{(1)}\, ,
\\
& & \nonumber \\
t
& = &
-\frac{2}{g_{4}}+\frac{i}{2}\varrho^{1/2}\, .
\end{eqnarray}

This solution is singular at $\varrho=0$ In particular, the imaginary part of
the scalar $t$ vanishes there. The underlying reason is that the
compactification circle's radius, measured by the KK scalar, shrinks to zero
at $\rho=0$. Asymptotically, the metric is conformal to that of
$\mathbb{R}\times S^{2}$, but it cannot be considered asymptotically-AdS$_{4}$
because the Weyl tensor diverges in this limit \cite{Ashtekar:1999jx}.  This
asymptotic behaviour is shared by all the asymptotically-AdS$_{5}$ solutions
written in the $k=1$ form, such as the Gutowski-Reall black hole.

Typically, some supersymmetry is always broken in the dimensional reduction of
AdS$_{5}$. This will happen if the 5-dimensional Killing vector depends on the
isometric coordinate $z$. To find whether this is the case and how much
supersymmetry can be preserved in 4 dimensions one has to solve explicitly the
Killing spinor equation which, for vanishing vector field strength, with our
choice of FI term, and setting $g=\sqrt{3}$, is given by

\begin{equation}
\delta_{\epsilon}\psi^{i}{}_{\mu}
=
\nabla_{\mu}\epsilon^{i}  
+\tfrac{i}{2} \sigma^{1\, i}{}_{j}\gamma_{\mu}\epsilon^{j} 
=
0\, ,
\end{equation}

\noindent
where $\sigma^{1}$ is the first Pauli matrix.

The $t$ component of this equation is

\begin{equation}
\left\{
\partial_{t}
+\tfrac{1}{4}\hat{J}_{mn}\gamma^{mn}+\tfrac{i}{2}\gamma^{0}\sigma^{1}
\right\}\epsilon
=
0\, ,  
\end{equation}

\noindent
and is solved by 

\begin{equation}
\epsilon =
e^{-\{\frac{1}{4}\hat{J}_{mn}\gamma^{mn}+\frac{i}{2}\gamma^{0}\sigma^{1}\}t}\,
\eta(\varrho,z,x^{1},x^{3})\, .  
\end{equation}

The $\varrho$ component of the Killing spinor equation reduces to the
following equation for the $t$-independent spinor $\eta$:

\begin{equation}
\left\{\partial_{\varrho}  
-H^{1/2}\tfrac{1}{2}\left(\gamma^{0\sharp} +i\gamma^{2}\sigma^{1} \right) \right\}
\eta
=
0\, ,
\end{equation}

\noindent
where $H^{-1}=\varrho(1+4\varrho)$, which is solved by 

\begin{equation}
\eta 
= 
e^{\int\! d\varrho H^{1/2}\frac{1}{2}\left(\gamma^{0\sharp} +i\gamma^{2}\sigma^{1} \right)}  
\xi(z,x^{1},x^{3})\, .
\end{equation}

The $z$ component, then, reduces to 

\begin{equation}
\left\{\partial_{z} +A\right\}\xi=0\, ,
\end{equation}

\noindent
where

\begin{equation}
  \begin{array}{rcl}
A  
& = & 
-e^{-B}\left\{ \tfrac{1}{8}\hat{J}_{mn}\gamma^{mn} 
+\left(2\varrho +H^{-1/2}\gamma^{0\sharp}\right)
\tfrac{1}{2}\left(\gamma^{\sharp 2}-i\gamma^{0}\sigma^{1}\right)
 \right\}
e^{B}\, ,
\\
& & \\
B
& = & 
{\displaystyle\int} d\varrho H^{1/2}\frac{1}{2}\left(\gamma^{0\sharp}
    +i\gamma^{2}\sigma^{1} \right)\, .
\end{array}
\end{equation}

\noindent
Since the Killing spinor equations are integrable, we know that $A$ is
$\varrho$-independent, but its actual value is important to determine whether
$\xi$, and hence $\epsilon$, is $z$-dependent or not.
 A long calculation gives 
$A=-\tfrac{1}{8}\hat{J}_{mn}\gamma^{mn}$ and

\begin{equation}
\xi = e^{\frac{1}{8}\hat{J}_{mn}\gamma^{mn}z}\, \zeta(x^{1},x^{3})\, .  
\end{equation}

The $z$-independent part of this spinor (and of the whole Killing spinor
$\epsilon$) is the one satisfying the projection

\begin{equation}
\tfrac{1}{2}\hat{J}_{mn}\gamma^{mn}\epsilon
= 
\gamma^{\sharp 2}\tfrac{1}{2}(1+\gamma^{\sharp 123})\epsilon
= 
\gamma^{\sharp 2}\, \tfrac{1}{2}(1+\gamma^{0})\epsilon
= 
0\, ,  
\end{equation}

\noindent
which is the condition generically satisfied by the timelike Killing spinors
of $\mathcal{N}=2,d=4$ theories. Most of the timelike supersymmetric solutions
of the minimal, gauged 5-dimensional supergravity must satisfy this condition
as well.

\subsection{Reduction of the Gutowski-Reall black hole}
\label{sec-reductionofGRBH}

The  Gutowski-Reall black hole is determined by 

\begin{eqnarray}
\hat{f} & = & \frac{\varrho}{\varrho +\frac{4\alpha^{2}-1}{4g^{2}}}\, ,
\hspace{1cm}
H^{-1} = \tfrac{4}{3}\varrho(3\alpha^{2} +g\varrho)\, ,  
\hspace{1cm}
W^{2} = \varrho H^{-1} \Phi_{(1)}\, ,
\nonumber \\
& & \\
\omega_{z} & = &
\frac{3(4\alpha^{2}-1)^{2}+24(4\alpha^{2}-1)g^{2} \varrho +32
  g^{4}\varrho^{2}}{16\sqrt{3}\varrho}\, ,
\hspace{1cm}
\omega = 0\, ,
\nonumber 
\end{eqnarray}

\noindent
and, according to the general rules, we get a 4-dimensional in which the two
1-form fields and the scalar field take non-trivial expressions. We are just
interested in the metric function and the KK scalar, which take the form

\begin{eqnarray}
k^{2}
& = &
\frac{ 3(4\alpha^{2}-1)^{3} +64 (4\alpha^{2}-1)^{2}(\alpha^{2}+2)g^{2}\varrho
+576 (4\alpha^{2}-1) g^{4}\varrho^{2} +768 g^{6} \varrho^{3}}
{48 g^{2}[(4\alpha^{2}-1)+ 4g^{2}\varrho]^{2}}\, , 
\\
& & \nonumber \\
e^{-2U}
& = &
\frac{3k [(4\alpha^{2}-1)+ 4g^{2}\varrho]}{16 g^{2}\varrho^{2} (3\alpha^{2}
  +g\varrho)}\, , 
\\
& & \nonumber \\
e^{-2U}W^{2}
& = &
k\left[\varrho +\frac{4\alpha^{2}-1}{4g^{2}}\right]\Phi_{(1)}\, .
\end{eqnarray}

In the $\varrho\rightarrow \infty$ limit the metric of this solution has the
same behaviour as that of the previous one. More interestingly, in the
$\varrho\rightarrow 0$ limit 

\begin{equation}
k^{2}\sim \frac{4\alpha^{2}-1}{16g^{2}}\equiv k^{2}_{\rm fix}\, ,
\hspace{1cm}
e^{-2U} \sim \frac{k^{3}_{\rm fix}}{\alpha^{2}}\frac{1}{\varrho^{2}}\, ,
\hspace{1cm}
e^{-2U}W^{2} \sim 4 k^{3}_{\rm fix}\, ,  
\end{equation}

\noindent
corresponding to an AdS$_{2}\times$S$^{2}$ near-horizon geometry in which the
two factor spaces have different radii. 

Thus, the Gutowski-Reall black hole reduces to a static, extremal,
4-dimensional black hole with exotic asymptotics.

\section{Conclusions}
\label{sec-conclusions}

In this paper we have shown how the metric ansatz of
Ref.~\cite{Chimento:2016run} simplifies the equations the determine the
timelike supersymmetric solutions of 5-dimensional minimal gauged supergravity
and allows one to find quite general families of interesting solutions such as
the black holes with non-compact horizons and the G\"odel-like solutions.

Our ansatz was inspired by the Gibbons-Hawking ansatz for the base space made
in Ref.~\cite{Gauntlett:2002nw} in the ungauged theory. However, there is a
very important difference between the gauged and ungauged cases (beyond the
K\"ahler and hyper-K\"ahler nature of the base spaces): in the ungauged case,
given a choice of base space, it is possible to construct many different
solutions which can be seen as ``excitations'' over the vacuum defined by the
choice: the choice of metric function $\hat{f}$ and of the harmonic function
$H$ that determines the Gibbons-Hawking metric are independent. In the gauged
case the situation is much more complicated because the base space is
different for each different solution: the functions $H$ and $W$ that define
the K\"ahler metric with one isometry depend on the metric function $\hat{f}$
and there is a different K\"ahler geometry for each solution. Of course, this
also happens for other K\"ahler metric ansatzs. With our ansatz this
dependence can be controlled more efficiently and it is possible to generate
systematically all the required K\"ahler solutions. The search for new
solutions is necessarily the search for new K\"ahler geometries or new forms
for the same K\"ahler geometries.

Another surprise we have found (in particular, in the study of the vacuum
solutions AdS$_{5}$) is the convenience (or even necessity) of using different
forms of the same base and how the coordinates of the base space (all
Euclidean in the base space) can have very different causal characters in the
full 5-dimensional metric.

The scope of our investigations was restricted to the simplest solutions with
an event horizon. These are black holes with only one independent angular
momentum. However, supersymmetric rotating black-hole solutions with more
independent angular momenta have also been constructed in
Ref.~\cite{Chong:2005da} and, associated to the general form of their base
space which can be adapted to our ansatz, we expect to find other families of
solutions. Furthermore, we would like to extend our results to matter-coupled
theories to reproduce and extend the results found in
Ref.~\cite{Kunduri:2006ek}.  Work in these directions is in progress
\cite{kn:CO}.

\section*{Acknowledgments}

TO would like to thank I.~Bena, M.~Dunajski, D.~Klemm, P.~Meessen, and
C.~Shahbazi for interesting and inspiring conversations. This work has been
supported in part by the Spanish Ministry of Science and Education grants
FPA2012-35043-C02-01 and FPA2015-66793-P, the Centro de Excelencia Severo
Ochoa Program grant SEV-2012-0249, and the Spanish Consolider-Ingenio 2010
program CPAN CSD2007-00042.  The work of SC has been supported by the Spanish
Ministry of Science and Education grant FPA2012-35043-C02-01. TO wishes to
thank M.M.~Fern\'andez for her permanent support.

\appendix
\section{3-d metrics}
\label{app-3dmetric}

Let us consider 3-dimensional Riemannian metrics of the form 

\begin{equation}
\label{eq:3dmetric}
d\overline{s}^{2}  
= 
\gamma_{\underline{i}\underline{j}}dx^{i}dx^{j}
=
(dx^{2})^{2} +W^{2}[(dx^{1})^{2}+(dx^{3})^{2}]\, ,
\end{equation}

\noindent
where $W$ depends on the three coordinates $x^{i}$, $i=1,2,3$ in an arbitrary
way. A convenient basis of Dreibeins is

\begin{equation}
\left\{
\begin{array}{rcl}
v^{1,3} & = & W dx^{1,3}\, ,\\
& & \\
v^{2} & = & dx^{2}\, ,\\
\end{array}
\right.
\hspace{2cm}
\left\{
\begin{array}{rcl}
v_{1,3} & = & W^{-1} \partial_{\underline{1},\underline{3}}\, ,\\
& & \\
v_{2} & = & \partial_{\underline{2}}\, .
\end{array}
\right.
\end{equation}

The non-vanishing components of the spin connection are

\begin{equation}
\overline{\omega}_{112}=\overline{\omega}_{332}= -\partial_{2}\log{W}\, ,
\hspace{1cm}
 \overline{\omega}_{113} = -\partial_{3}\log{W}\, , 
\hspace{1cm}
 \overline{\omega}_{331} = -\partial_{1}\log{W}\, ,
\end{equation}

\noindent
and those of the Riemann curvature tensor are

\begin{equation}
  \begin{array}{rclrcl}
\overline{R}_{1212}
& = & 
\overline{R}_{2323}
=
W^{-1}\partial^{2}_{\underline{2}}W\, ,
&
\overline{R}_{1213}
& = & 
W^{-1}\partial_{\underline{2}}\partial_{\underline{3}}\log{W}\, ,
\\
& & & & & \\
\overline{R}_{1313}
& = & 
W^{-2}\left( \partial^{2}_{\underline{1}}+\partial^{2}_{\underline{3}}\right)
\log{W}
+\left(\partial_{\underline{2}}W \right)^{2}\, ,
&
\overline{R}_{1323}
& = & 
W^{-1}\partial_{\underline{2}}\partial_{\underline{1}}\log{W}\, ,
\end{array}
\end{equation}

\noindent
those of the Ricci tensor are

\begin{equation}
\begin{array}{rclrcl}
\overline{R}_{11}
& = & 
\tfrac{1}{2}
W^{-2}\left( \partial^{2}_{\underline{1}}+\partial^{2}_{\underline{3}}
+W^{2}\partial^{2}_{\underline{2}}\right)
\log{W^{2}} 
+\tfrac{1}{2}\left(\partial_{\underline{2}}\log{W^{2}} \right)^{2}\, ,    
& 
\overline{R}_{22}
& = & 
W^{-1}\partial^{2}_{\underline{2}}W\, ,
\\
& & & &  & \\
\overline{R}_{12}
& = & 
\tfrac{1}{2}
W^{-1}\partial_{\underline{1}}\partial_{\underline{2}}\log{W^{2}}\, ,
& 
\overline{R}_{23}
& = & 
\tfrac{1}{2}
W^{-1}\partial_{\underline{3}}\partial_{\underline{2}}\log{W^{2}}\, ,
\\
& & & & & \\
\overline{R}_{33}
& = &
\overline{R}_{11}\, .
& & & \\
\end{array}
\end{equation}

The Ricci scalar is given by

\begin{equation}
\overline{R}
=
W^{-2}\left( \partial^{2}_{\underline{1}}+\partial^{2}_{\underline{3}}
+2W^{2}\partial^{2}_{\underline{2}}\right)
\log{W^{2}} 
+2\left(\partial_{\underline{2}}\log{W^{2}} \right)^{2}\, .      
\end{equation}

\section{4-d Euclidean metrics with one isometry}
\label{app-4dmetric}

Any 4-dimensional Euclidean metric admitting one isometry can be written in
the form 

\begin{equation}
\label{eq:4dmetric}
d\hat{s}^{2} 
= 
H^{-1}(dz+\chi)^{2} 
+
H\gamma_{\underline{i}\underline{j}}dx^{i}dx^{j}\, , 
\end{equation}

\noindent
where $z=x^{\sharp}$ is the coordinate adapted to the isometry and where the
3-dimensional function $H$, the 1-form $\chi=\chi_{\underline{i}}dx^{i}$ and
the metric $\gamma_{\underline{i}\underline{j}}dx^{i}dx^{j}$, $i,j=1,2,3$ are
$z$-independent and orthogonal to the Killing vector
$k^{\underline{m}}=\delta_{z}{}^{\underline{m}}$. We denote the world indices
by $\{\underline{m}\}=\{z,\underline{i}\}$ and the flat indices by $\{ m
\}=\{\sharp,i\}$. We will denote 3-dimensional structures (connection,
curvature etc.) by an overline, as in the previous appendix. For the moment,
the 3-dimensional structures will be completely general and only later on we
will assume the 3-dimensional metric to have the form Eq.~(\ref{eq:3dmetric})
and $H$, $\chi$ and $W$ to be related by th $W$-deformed monopole equation
(\ref{eq:constraintijcurved}) which holds when the 4-dimensional metric above
is a K\"ahler metric with respect to the complex structure
Eq.~(\ref{eq:J_matrix}).

A convenient basis of Vierbeins is 

\begin{equation}
\label{eq:4dvierbein}
\left\{
\begin{array}{rcl}
\hat{V}^{\sharp} & = & H^{-1/2}(dz+\chi)\, ,\\
& & \\
\hat{V}^{i} & = & H^{1/2}v^{i}\, ,\\
\end{array}
\right.
\hspace{2cm}
\left\{
\begin{array}{rcl}
\hat{V}_{\sharp} & = & H^{1/2} \partial_{z}\, ,\\
& & \\
\hat{V}_{i} & = & H^{-1/2}(\partial_{i} -\chi_{i}\partial_{z})\, ,
\end{array}
\right.
\end{equation}

\noindent
where $v^{i}= v^{i}{}_{\underline{j}}dx^{j}$ are Dreibeins of the metric
$\gamma_{\underline{i}\underline{j}}$, $\partial_{i}\equiv
v_{i}{}^{\underline{j}}\partial_{\underline{j}}$ and $\chi_{i}\equiv
v_{i}{}^{\underline{j}}\chi_{\underline{j}}$.

The non-vanishing components of the spin connection 1-form, defined through
the structure equation $\mathcal{D}e^{m}\equiv de^{m}-\varpi^{m}{}_{n}\wedge
e^{n}=0$ are

\begin{equation}
\label{eq:4d_spin_conn}
\begin{array}{rclrcl}
\varpi_{\sharp \sharp i} 
& = & 
\tfrac{1}{2} H^{-3/2}\partial_{i}H\, ,
\hspace{1.5cm}
&
\varpi_{\sharp ij} 
& = & 
\tfrac{1}{2} H^{-3/2} (d\chi)_{ij}\, ,
\\
& & & & & \\
\varpi_{i\sharp j} 
& = & 
\varpi_{\sharp ij}\, , 
&
\varpi_{kij} 
& = & 
H^{-1/2}\overline{\omega}_{kij} 
+H^{-3/2}\partial_{[i}H\delta_{j]k}\, ,
\end{array}
\end{equation}

\noindent
where $(d\chi)_{ij} = 2v_{i}{}^{\underline{k}}
v_{j}{}^{\underline{l}}\partial_{[\underline{j}}\chi_{\underline{l}]}$ and
$\overline{\omega}_{kij}$ is the 3-dimensional connection defined by
$\overline{\mathcal{D}} v^{i}=dv^{i}-\overline{\omega}^{i}{}_{j}\wedge v^{j}=0$.

Those of the curvature 2-form, defined through $\hat{R}^{m}{}_{n}\equiv
d\varpi^{m}{}_{n}-\varpi^{m}{}_{p}\wedge \varpi^{p}{}_{n}$, are  

\begin{equation}
  \begin{array}{rcl}
\hat{R}_{\sharp i\sharp j} 
& = & 
-\tfrac{1}{2} H^{-2} \overline{\nabla}_{j}\partial_{i}H
+\tfrac{1}{4} H^{-3}
\left[
5\partial_{i}H\partial_{j}H -\delta_{ij}(\partial H)^{2}
-(d\chi)_{jl}(d\chi)_{il}
\right]\, ,
\\
& & \\  
\hat{R}_{kj\sharp i}
& = &
\overline{\nabla}_{k}[H^{-2}(d\chi)_{ji}]
+\tfrac{1}{2}H^{-3}
\left[
2\partial_{[k}H(d\chi)_{j]i}
+\partial_{l}H(d\chi)_{l[k}\delta_{j]i}
\right]\, ,
\\
& & \\
\hat{R}_{klij}
& = &
H^{-1}
\left\{
\overline{R}_{klij}
+2H^{-1}\overline{\nabla}_{[k}\partial^{[i}H\delta^{j]}{}_{l]}
+3H^{-2} \partial_{[i}H \delta_{j][k}\partial_{l]}H
\right.
\\
& & \\
& & 
\left.
+\tfrac{1}{2}H^{-2}(\partial H)^{2}\delta_{ij\, , \, kl}
+\tfrac{1}{2}H^{-2}
\left[
(d\chi)_{ij}(d\chi)_{kl} -(d\chi)_{i[k}(d\chi)_{l]j}
\right]
\right\}\, .
\end{array}
\end{equation}

The components of the Ricci tensor are

\begin{equation}
\begin{array}{rcl}
\hat{R}_{\sharp \sharp }
& = &
-\tfrac{1}{2}H^{-2}\overline{\nabla}^{2}H    
+\tfrac{1}{2}H^{-3}(\partial H)^{2} 
-\tfrac{1}{4}H^{-3}(d\chi)^{2}\, ,
\\
& & \\
\hat{R}_{\sharp i}
& = &
\tfrac{1}{2}\overline{\nabla}_{j}\left[H^{-2}(d\chi)_{ji}\right]\, ,
\\
& & \\
\hat{R}_{ij}
& = &
H^{-1}\overline{R}_{ij} +\tfrac{1}{2}\delta_{ij}H^{-2}\overline{\nabla}^{2}H
+\tfrac{1}{2}
H^{-3}\left[
\partial_{i}H\partial_{j}H -\delta_{ij}(\partial H)^{2}
+(d\chi)_{ik}(d\chi)_{jk}
\right]\, ,
\end{array}
\end{equation}

\noindent
and the Ricci scalar is given by

\begin{equation}
\hat{R} = H^{-1}\overline{R} +H^{-2}\overline{\nabla}^{2}H  
-\tfrac{1}{2}
H^{-3}\left[
(\partial H)^{2}
-\tfrac{1}{2}(d\chi)^{2}
\right]\, .
\end{equation}

Observe that if the conditions

\begin{equation}
\overline{R}_{ij}=0\, ,
\hspace{1cm}
(d\chi)_{ij} =\varepsilon_{ijk}\partial_{l}H\, ,  
\end{equation}

\noindent
are satisfied the metric Eq.~(\ref{eq:4dmetric}) is a Gibbons-Hawking metric
(a hyperK\"ahler metric admitting a triholomorphic isometry)
\cite{Gibbons:1979zt,Gibbons:1979xm} and it is Ricci-flat. If the metric is
K\"ahler with respect to the complex structure Eq.~(\ref{eq:J_matrix}) so that
the 3-dimensional metric has the form Eq.~(\ref{eq:3dmetric}) and $H$, $\chi$
and $W$ are related by the $W$-deformed monopole equation
(\ref{eq:constraintijcurved}), then we can use the results in
Appendix~\ref{app-3dmetric} to find that the non-vanishing components of the
Ricci tensor are given by

\begin{equation}
\begin{array}{rclcl}
\hat{R}_{\sharp \sharp}  
& = &
\hat{R}_{22}
& = &
\tfrac{1}{2}\partial_{\underline{2}}\left(H^{-1}\partial_{\underline{2}}\log{W^{2}}
\right)\, ,
\\
& & & & \\
\hat{R}_{11}
& = &
\hat{R}_{33}
& = & 
\tfrac{1}{2}H^{-1}W^{-2}
\left(\partial^{2}_{\underline{1}}+\partial^{2}_{\underline{3}} \right)
\log{W^{2}}
+\tfrac{1}{2}H^{-1}\left(\partial_{\underline{2}}\log{W^{2}} \right)^{2}
\\
& & & & \\
& & & & 
+\tfrac{1}{2}H^{-2} \partial_{\underline{2}}H\partial_{\underline{2}}\log{W^{2}}\, ,
\\
& & & & \\
\hat{R}_{01}
& = &
\hat{R}_{23}
& = &
-\tfrac{1}{2}H^{-2}W^{-1} \partial_{\underline{3}}H\partial_{\underline{2}}\log{W^{2}} 
+\tfrac{1}{2}H^{-1}W^{-1} \partial_{\underline{3}}\partial_{\underline{2}}\log{W^{2}}\, ,
\\
& & & & \\
\hat{R}_{03}
& = &
-\hat{R}_{12}
& = &
\tfrac{1}{2}H^{-2}W^{-1} \partial_{\underline{1}}H\partial_{\underline{2}}\log{W^{2}} 
-\tfrac{1}{2}H^{-1}W^{-1} \partial_{\underline{1}}\partial_{\underline{2}}\log{W^{2}}\, .
\end{array}
\end{equation}

\noindent
Exactly the same result is obtained by using Eq.~(\ref{eq:Rmn}).

Finally, the Ricci scalar is given by 

\begin{equation}
\begin{array}{rcl}
\hat{R} 
& = &
\hat{\nabla}^{2}\log{W^{2}} 
=   
H^{-1}\overline{\nabla}^{2}\log{W^{2}} 
\\
& & \\
& = &
H^{-1}W^{-2} 
\left\{
\left(\partial^{2}_{\underline{1}}+\partial^{2}_{\underline{3}} \right)
\log{W^{2}}
+\partial_{\underline{2}}\left(W^{2}\partial_{\underline{2}}\log{W^{2}} \right)
\right\}\, .   
\end{array}
\end{equation}

\section{5-d metrics}
\label{app-5dmetric}

Let us consider the time-independent 5-dimensional Lorentzian
conformastationary metric

\begin{equation}
ds^{2} = \hat{f}^{2}\left(dt+\hat{\omega}\right)^{2}
-\hat{f}^{-1}h_{\underline{m}\underline{n}} dx^{m}dx^{n}\, ,
\hspace{2cm}
m,n=\sharp,1,2,3\, .
\end{equation}

\noindent
The function $\hat{f}$ and the 1-form
$\hat{\omega}=\hat{\omega}_{\underline{m}}dx^{m}$ can be understood as objects
living in the 4-dimensional Euclidean metric
$h_{\underline{m}\underline{n}}$. We will denote this kind of objects with
hats.

We choose the Vielbein basis

\begin{equation}
\begin{array}{rclrcl}
e^{0}
& = &
\hat{f}(dt+\hat{\omega})\, ,
\hspace{1.5cm}
&
e_{0}
& = &
\hat{f}^{-1}\partial_{t}\, ,
\\
& & & & & \\
e^{m}
& = &
\hat{f}^{-1/2}\hat{V}^{m}\, ,
&
e_{m}
& = &
\hat{f}^{1/2}
(\partial_{m} -\hat{\omega}_{m}\partial_{t})\, .     
\end{array}
\end{equation}

\noindent
where the $\hat{V}_{\underline{m}}{}^{p}$s are a \index{Vierbein} Vierbein for
the 4-dimensional Euclidean metric $h_{\underline{m}\underline{n}}$ and, just
as we did with the 3- and 4-dimensional metrics studied before, all the
objects in the r.h.s.~of all the equations refer to the 4-dimensional metric
$h_{\underline{m}\underline{n}}$ and the Vierbein basis $\hat{V}^{p}$
($\partial_{m}=V_{m}{}^{\underline{p}}\partial_{\underline{p}}$).


With this choice of Vielbein, the non-vanishing components of the spin
connection are

\begin{equation}
\label{eq:conformaspincon}
\begin{array}{rclrcl}
\omega_{00m} 
& = &
-2\partial_{m}\hat{f}^{1/2}\, ,
\hspace{1.5cm}
&
\omega_{0mn}
& = &
\tfrac{1}{2}\hat{f}^{2}\left(d\hat{\omega}\right)_{mn}\, ,
\\
& & & & & \\
\omega_{m0n}
& = &
\tfrac{1}{2}\hat{f}^{2}\left(d\hat{\omega}\right)_{mn}\, ,
&
\omega_{mnp}
& = &
-\hat{f}^{1/2}\varpi_{mnp}-2\delta_{m[n}\partial_{p]}\hat{f}^{1/2}\, ,
\end{array}
\end{equation}

\noindent
where we are denoting by $\varpi_{mnp}$ the 4-dimensional spin connection.

The non-vanishing components of the Ricci tensor are

\begin{equation}
 \begin{array}{rcl}
R_{00} 
& = & 
-\hat{\nabla}^{2}\hat{f} +\hat{f}^{-1}(\partial \hat{f})^{2}
-\tfrac{1}{4}\hat{f}^{4}(d\hat{\omega})^{2}\, ,
\\
& & \\
R_{0m} 
& = & 
-\tfrac{1}{2}\hat{f}^{-1/2}\hat{\nabla}_{n}[\hat{f}^{3}(d\hat{\omega})_{nm}]\,
,
\\
& & \\
R_{mn} 
& = & 
\hat{f}\hat{R}_{mn} -\frac{1}{2}(d\hat{\omega})_{mp}(d\hat{\omega})_{np}
+\tfrac{3}{2}\hat{f}^{-1}\partial_{m}\hat{f}\partial_{n}\hat{f} 
-\tfrac{1}{2}\delta_{mn}[\nabla^{2}\hat{f} -\hat{f}^{-1}(\partial
\hat{f})^{2}]\, ,
\\
\end{array}
\end{equation}

\noindent
and the Ricci scalar is given by

\begin{equation}
R
= 
-\hat{f}\hat{R} 
+\tfrac{1}{4}(d\hat{\omega})^{2} +\hat{\nabla}^{2}\hat{f} 
-\tfrac{5}{2}\hat{f}^{-1}(\partial \hat{f})^{2}\, .
\end{equation}

\section{AdS$_{5}$}
\label{app-ads5}

It is well known that (the unit radius) AdS$_{5}$ can be embedded in
$\mathbb{R}^{2,4}$ or equivalently in $\mathbb{C}^{1,2}$ as the set of points
satisfying

\begin{equation}
 Z^{0}Z^{*\, 0}-Z^{i}Z^{*\, i}=1\, ,
\hspace{1cm}
i=1,2
\end{equation}

\noindent
with its metric being induced from the ambient metric

\begin{equation}
ds^{2}=  dZ^{0}dZ^{*\, 0}-dZ^{i}dZ^{*\, i}\, .
\end{equation}

Setting $Z^{0} = |Z^{0}|e^{it}$, $Z^{i} = Z^{0} \zeta^{i}$ we can solve for
$Z^{0}$ in terms of $t$ and $\zeta^{i}$

\begin{equation}
 |Z^{0}|^{-2} = 1-\zeta^{i}\zeta^{*\, i^{*}}\, ,
\end{equation}

\noindent
and the induced metric takes the form

\begin{equation}
ds^{2} 
= \left( dt+\mathcal{Q} \right)^{2}
-2\mathcal{G}_{ij^{*}}d\zeta^{i} d\zeta^{*\, j^{*}}\, ,
\end{equation}

\noindent
where

\begin{equation}
2\mathcal{G}_{ij^{*}}
=
\frac{ \delta_{ij^{*}}}{1-\zeta^{k}\zeta^{*\, k^{*}}}
+
  \frac{\zeta^{*\, i^{*}}\zeta^{j}}{(1-\zeta^{k}\zeta^{*\, k^{*}})^{2}}\, ,  
\end{equation}

\noindent
is the metric of the K\"ahler space
$\mathbb{\overline{CP}}^{2}=$SU$(1,2)/$U$(2)$ and

\begin{equation}
\mathcal{Q} 
= 
\tfrac{i}{2}\frac{\zeta^{*\, i^{*}}d\zeta^{i}-\zeta^{i} d\zeta^{*\,i^{*}}}
{1-\zeta^{i}\zeta^{*\, i^{*}}}\, ,
\end{equation}

\noindent
is its corresponding K\"ahler 1-form connection. The K\"ahler 2-form is given
by

\begin{equation}
\mathcal{J}_{ij^{*}} 
= 
\partial_{i}\mathcal{Q}_{j^{*}}
- \partial_{j^{*}}\mathcal{Q}_{i}
= 
2i\mathcal{G}_{ij^{*}}\, .
\end{equation}

This form of the metric makes manifest that AdS$_{5}$ can be seen as a U$(1)$
fibration over the K\"ahler manifold $\mathbb{\overline{CP}}^{2}$.  As shown
in Ref.~\cite{Gutowski:2004ez} this is the only base space that can be used to
construct AdS$_{5}$ as a supersymmetric solution of minimal gauged
5-dimensional supergravity. There are different ways of writing
$\mathbb{\overline{CP}}^{2}$ in the canonical form
Eqs.~(\ref{eq:final_metric}) and (\ref{eq:constraintijcurved}), associated to
the different holomorphic Killing vectors of the manifold which, being the
symmetric space SU$(2,1)/$U$(2)$, are 8. We are not going to explore all of
them here. We will content ourselves with those in which the metric contains
the metric of a 2-dimensional space of constant curvature $k$ that we will
denote by $d\Omega_{(2,k)}$, where $k=1,0,-1$ for, respectively, S$^{2}$,
$\mathbb{E}^{2}$ or $\mathbb{H}_{2}$.

\subsection{$k=1$}

In the $k=1$ case we can use the real coordinates

\begin{equation}
\label{eq:cp2_k1_coords}
\zeta^{1} 
= \tanh{\rho}\cos{\tfrac{\theta}{2}}\, e^{-\frac{i}{2}(z+\varphi)}\, ,
\hspace{1.5cm} 
\zeta^{2} 
= 
\tanh{\rho}\sin{\tfrac{\theta}{2}}\,e^{-\frac{i}{2}(z-\varphi)}\, ,
\end{equation}

\noindent
for which the metric of $\mathbb{\overline{CP}}^{2}$ and the K\"ahler 1-form
connection are given by 

\begin{equation}
\begin{array}{rcl}
ds_{\mathbb{\overline{CP}}^{2}}^{2}
& = & 
d\rho^{2} 
+\tfrac{1}{4} \sinh^{2}{\rho}\cosh^{2}{\rho}
\left( dz+\cos{\theta} d\varphi\right)^{2}
+\tfrac{1}{4} \sinh^{2}{\rho} d\Omega^{2}_{(2,1)}\, ,
\\
& & \\
\mathcal{Q}_{\mathbb{\overline{CP}}^{2}}
& = & 
\tfrac{1}{2} \sinh^{2}{\rho}
\left( dz+\cos{\theta}\, d\varphi \right)\, .
\end{array}
\end{equation}

\noindent
where

\begin{equation}
d\Omega^{2}_{(2,1)}
=
d\theta^{2}+\sin^{2}{\theta}\, d\varphi^{2}\, ,  
\end{equation}

\noindent
is the metric of $S^{2}$.

The metric for the four-dimensional base space can be cast in the form
Eq.~(\ref{eq:final_metric}) by defining the new coordinates

\begin{equation}
x^{1} = \tan{\tfrac{\theta}{2}}\cos{\varphi}\, ,
\hspace{1cm} 
x^{2} = \tfrac{1}{4}\sinh^{2}{\rho}\, ,
\hspace{1cm} 
x^{3} = \tan{\tfrac{\theta}{2}}\sin{\varphi}\, ,
\end{equation}

\noindent
so that the functions $H,W$ and 1-form $\chi_{(1)}$ that define it are given
by\footnote{These functions have been determined for
  $\mathbb{\overline{CP}}^{2}$ in Ref.~\cite{Dunajski:2013qc}.}

\begin{equation}
\begin{array}{rcl}
H^{-1} 
& = & 
x^{2}( 1+4x^{2})\, ,
\\
& & \\
W^{2} 
& = &
{\displaystyle
\frac{4x^{2}}{H[1+(x^{1})^{2}+(x^{3})^{2}]^{2}}\, ,
}
\\
& & \\
\chi
& = & 
\chi_{(1)}
\equiv
{\displaystyle
\frac{[1-(x^{1})^{2}-(x^{3})^{2}]}{[1+(x^{1})^{2}+(x^{3})^{2}]}
\frac{x^{1}dx^{3}-x^{3}dx^{1}}{(x^{1})^{2}+(x^{3})^{2}}\, ,
}
\end{array}
\end{equation}

\noindent
and\footnote{The 1-form $\chi_{(1)}$ is defined up to a total derivative that
  can be absorbed in a redefinition of the coordinate $z$. The expression
  given above for $\chi_{(1)}$ is exactly the one that appears in the
  metric. A simpler expression is
\begin{equation}
\chi_{(1)} 
=
\frac{x^{3}dx^{1}-x^{1}dx^{3}}{1+(x^{1})^{2}+(x^{3})^{2}}\, .
\end{equation}
}

\begin{equation}
d\chi_{(1)} = -\frac{4}{[1+(x^{1})^{2}+(x^{3})^{2}]^{2}}dx^{1}\wedge dx^{3}\,.
\end{equation}

\noindent
From these expressions it is trivial to verify that the constraints
\eqref{eq:constraintijcurved} are satisfied.

Using the parametrization (\ref{eq:cp2_k1_coords}) for $\mathbb{\overline{CP}}^{2}$ we find the following
line element of AdS$_{5}$

\begin{multline}
ds^{2} 
= 
\left[dt+\tfrac{1}{2} \sinh^{2}{\rho}
\left( dz+\cos{\theta}\, d\varphi \right)\right]^{2}
\\ \\
-d\rho^{2} 
-\tfrac{1}{4} \sinh^{2}{\rho}\cosh^{2}{\rho}\left( dz+\cos{\theta} d\varphi
\right)^{2}
-\tfrac{1}{4} \sinh^{2}{\rho} d\Omega^{2}_{(2,1)}\, .
\end{multline}

The off-diagonal components can be eliminated by redefining the angular
coordinate $z= \psi +2t$:

\begin{equation}
\label{eq:AdS5diagonalmetric}
ds^{2} 
= 
\cosh^{2}{\rho}\, dt^{2} -d\rho^{2} -\sinh^{2}{\rho}\, d\Omega^{2}_{(3,1)}\, ,
\end{equation}

\noindent
where 

\begin{equation}
d\Omega^{2}_{(3,1)} 
= 
\tfrac{1}{4}
\left[ 
\left( 
d\psi^{\prime}+\cos{\theta}\, d\varphi \right)^{2}+d\Omega^{2}_{(2,1)}
\right]
\end{equation}

\noindent
is the metric of the round 3-sphere of unit radius. This is one of the
standard expressions for the metric of AdS$_{5}$ in global coordinates. The
coordinates used in the supersymmetric form (rotating frame $\psi\rightarrow
z$) also cover the whole AdS$_{5}$ spacetime.

Redefining the radial coordinate $r=\sinh{\rho}$ the metric takes the standard
form

\begin{equation}
\label{eq:standardAdS5metric}
ds^{2} = (1+r^{2})dt^{2} -\frac{dr^{2}}{1+r^{2}} -r^{2}d\Omega^{2}_{(3)}\, .  
\end{equation}

Using the results in the previous appendices one finds that the Ricci tensor
of this metric is $R_{ab} = -4\eta_{ab}$. In order to get a metric satisfying
$R_{ab}=\Lambda \eta_{ab}$ (for $\Lambda<0$) where $\Lambda$ is the
cosmological constant as defined in footnote~\ref{foot:cosmocon} we just have
to multiply the whole metric by $4/|\Lambda|$.  In particular, if we multiply
the AdS$_{5}$ metric in Eq.~(\ref{eq:AdS5diagonalmetric}) by that factor and
make the coordinate redefinitions $r= \sqrt{4/|\Lambda|} \sinh{\rho}$ and
$t^{\prime}= \sqrt{4/|\Lambda|}\, t$ we get, instead of
Eq.~(\ref{eq:standardAdS5metric})

\begin{equation}
\label{eq:standardAdS5metriclambda}  
ds^{2} = \left(1+\frac{|\Lambda|}{4}r^{2}\right)dt^{\prime\, 2} 
-\left(1+\frac{|\Lambda|}{4}r^{2}\right)^{-1}dr^{2}
-r^{2}d\Omega^{2}_{(3)}\, .  
\end{equation}

\subsection{$k=0$}

In the $k=0$ case the real coordinates one has to use for
$\mathbb{\overline{CP}}^{2}$ are essentially the ones customarily used to
parametrize the universal hypermultiplet\footnote{See, for instance,
  Ref.~\cite{Ketov:2001gq} and references therein}:

\begin{equation}
\zeta^{1} = \frac{1-S}{1+S}\, ,
\hspace{1cm}
\zeta^{2} = \frac{2C}{1+S}\, ,  
\hspace{1cm}
\mbox{with}
\hspace{1cm}
\left\{
\begin{array}{rcl}
S & = & {\displaystyle\frac{1}{x^{2}}} +4iz +CC^{*}\, ,\\
& & \\
C & = & 2(x^{1}+ix^{3})\, .\\
\end{array}
\right.
\end{equation}

In terms of these coordinates, the metric of $\mathbb{\overline{CP}}^{2}$ and
the K\"ahler 1-form connection are given by

\begin{equation}
\begin{array}{rcl}
ds_{\mathbb{\overline{CP}}^{2}}^{2}
& = & 
{\displaystyle\frac{(dx^{2})^{2}}{4(x^{2})^{2}}}
+4(x^{2})^{2}\left[ dz+2(x^{3}dx^{1}-x^{1}dx^{3})\right]^{2}
+x^{2}d\Omega^{2}_{(2,0)}\, ,
\\
& & \\
\mathcal{Q}_{\mathbb{\overline{CP}}^{2}}
& = & 
2x^{2}
\left[ dz+2(x^{3}dx^{1}-x^{1}dx^{3})\right]\, .
\end{array}
\end{equation}

\noindent
where

\begin{equation}
d\Omega^{2}_{(2,0)}
=
4[(dx^{1})^{2}+(dx^{3})^{2}]\, ,  
\end{equation}

\noindent
is the metric of $\mathbb{E}^{2}$ with a convenient normalization.

This metric is already in the form Eq.~(\ref{eq:final_metric}) and so that the
functions $H,W$ and 1-form $\chi$ that define it are given by\footnote{These
  functions have been determined for $\mathbb{\overline{CP}}^{2}$ with $k=1$
  in Ref.~\cite{Dunajski:2013qc}.}

\begin{equation}
\begin{array}{rcl}
H^{-1} 
& = & 
4(x^{2})^{2}\, ,
\\
& & \\
W^{2} 
& = &
{\displaystyle
\frac{x^{2}}{H}\Phi_{(0)}\, ,
}
\\
& & \\
\chi
& = & 
\chi_{(0)} 
\equiv
2(x^{3}dx^{1}-x^{1}dx^{3})\, .
\end{array}
\end{equation}

Using these coordinates for $\mathbb{\overline{CP}}^{2}$ we find the following
line element of AdS$_{5}$

\begin{multline}
ds^{2} 
= 
\left\{dt+2x^{2}
\left[ dz+2(x^{3}dx^{1}-x^{1}dx^{3}) \right]\right\}^{2}
\\ \\
-{\displaystyle\frac{(dx^{2})^{2}}{4(x^{2})^{2}}}
-4(x^{2})^{2}\left[ dz+2(x^{3}dx^{1}-x^{1}dx^{3})\right]^{2}
-x^{2}d\Omega^{2}_{(2,0)}\, .
\end{multline}

In this case we cannot eliminate the off-diagonal components of the metric
with a simple coordinate transformation.

\subsection{$k=-1$}

In the $k=-1$ case  we can use the real coordinates

\begin{equation}
\label{eq:cp2_k-1_coords}
\zeta^{1} 
= \tanh{(\theta/2)}\, e^{i\varphi}\, ,
\hspace{1.5cm} 
\zeta^{2} 
= 
\frac{\tanh{\rho}}{\cosh{(\theta/2)}}\,e^{-\frac{i}{2}(z-\varphi)}\, ,
\end{equation}

\noindent
for which the metric of $\mathbb{\overline{CP}}^{2}$ and the K\"ahler 1-form
connection are given by 

\begin{equation}
\begin{array}{rcl}
ds_{\mathbb{\overline{CP}}^{2}}^{2}
& = & 
d\rho^{2} 
+\tfrac{1}{4} \sinh^{2}{\rho}\cosh^{2}{\rho}
\left( dz-\cosh{\theta} d\varphi\right)^{2}
+\tfrac{1}{4} \cosh^{2}{\rho}\, d\Omega^{2}_{(2,-1)}\, ,
\\
& & \\
\mathcal{Q}_{\mathbb{\overline{CP}}^{2}}
& = & 
\tfrac{1}{2} \cosh^{2}{\rho}
\left( dz-\cosh{\theta}\, d\varphi \right)\, .
\end{array}
\end{equation}

\noindent
where

\begin{equation}
d\Omega^{2}_{(2,-1)}
=
d\theta^{2}+\sinh^{2}{\theta}\, d\varphi^{2}\, ,  
\end{equation}

\noindent
is the metric of the $\mathbb{H}_{2}$. Observe that now $\theta$ is
a non-compact coordinate. 

To cast the above metric in the form Eq.~(\ref{eq:final_metric}) we define
\begin{equation}
x^{1} = \tanh{\tfrac{\theta}{2}}\cos{\varphi}\, ,
\hspace{1cm} 
x^{2} = \tfrac{1}{4}\cosh^{2}{\rho}\, ,
\hspace{1cm} 
x^{3} = \tanh{\tfrac{\theta}{2}}\sin{\varphi}\, .
\end{equation}
Then, the functions $H,W$ and 1-form $\chi$
that define it are given by\footnote{Again, the expression given above for
  $\chi_{(-1)}$ is exactly the one that appears in the metric. A simpler
  expression is
\begin{equation}
\chi_{(-1)} 
=
\frac{x^{3}dx^{1}-x^{1}dx^{3}}{1-(x^{1})^{2}-(x^{3})^{2}}\, .
\end{equation}
}

\begin{equation}
\begin{array}{rcl}
H^{-1} 
& = & 
x^{2}( -1+4x^{2})\, ,
\\
& & \\
W^{2} 
& = &
{\displaystyle
\frac{4x^{2}}{H[1-(x^{1})^{2}-(x^{3})^{2}]^{2}}\, ,
}
\\
& & \\
\chi
& = & 
\chi_{(-1)}
\equiv
{\displaystyle
\frac{[1+(x^{1})^{2}+(x^{3})^{2}]}{[1-(x^{1})^{2}-(x^{3})^{2}]}
\frac{x^{1}dx^{3}-x^{3}dx^{1}}{(x^{1})^{2}+(x^{3})^{2}}\, .
}
\end{array}
\end{equation}

The line element for AdS$_{5}$ corresponding to the choice of coordinates (\ref{eq:cp2_k-1_coords}) is 

\begin{multline}
ds^{2} 
= 
\left[dt+\tfrac{1}{2} \cosh^{2}{\rho}
\left( dz-\cosh{\theta}\, d\varphi \right) \right]^{2}
\\ \\
-d\rho^{2} 
-\tfrac{1}{4} \sinh^{2}{\rho}\cosh^{2}{\rho}
\left( dz-\cosh{\theta} d\varphi\right)^{2}
-\tfrac{1}{4} \cosh^{2}{\rho}\, d\Omega^{2}_{(2,-1)}\, .
\end{multline}

Observe that, if we eliminate the $dtdz$ terms in the $k=-1$ metric
using the same trick as in the $k=1$ case, namely shifting the $z$ coordinate
$z=\psi-2t$, we get the metric 

\begin{equation}
ds^{2}
=
-\sinh^{2}{\rho}dt^{2} +\tfrac{1}{4}\cosh^{2}{\rho}\left( d\psi+\cosh{\theta}d\varphi\right)^{2}
-d\rho^{2} -\tfrac{1}{4} \cosh^{2}{\rho}\, d\Omega^{2}_{(2,-1)}\, ,
\end{equation}

\noindent
in which $t$ and $\psi$ have interchanged their r\^oles.

The functions corresponding to the three different canonical metrics for
$\mathbb{\overline{CP}}^{2}$ can be written in a unified form:

\begin{equation}
\begin{array}{rcl}
H^{-1} 
& = & 
x^{2}(k+4x^{2})\, ,
\\
& & \\
W^{2} 
& = &
{\displaystyle
\frac{x^{2}}{H}\Phi_{(k)}\, ,
}
\\
& & \\
\chi
& = & 
\chi_{(k)}
\end{array}
\end{equation}

\noindent
with 

\begin{equation}
\label{eq:Phidef}
  \begin{array}{rcl}
d\Omega^{2}_{(2,k)} 
& = &
{\displaystyle\frac{4  [(dx^{1})^{2}+(dx^{3})^{2}]}{\{1+k[(x^{1})^{2}+(x^{3})^{2}]\}^{2}}
}
\equiv \Phi_{(k)}(x^{1},x^{3})   [(dx^{1})^{2}+(dx^{3})^{2}]\, ,
\\
& & \\
\chi_{(k)}
& = &
{\displaystyle
\frac{2[x^{3}dx^{1}-x^{1}dx^{3}]}{1+k[(x^{1})^{2}+(x^{3})^{2}]}
} \, .
\end{array}
\end{equation}

Then, the metric of AdS$_{5}$ in the supersymmetric canonical form is given by 

\begin{equation}
ds^{2}
=\left[ dt+ 2x^{2}  (dz+\chi_{(k)})\right]^{2}
-x^{2}(k+ 4x^{2})(dz+\chi_{(k)})^{2}
-\frac{(dx^{2})^{2}}{x^{2}(k+ 4x^{2})}
-x^{2} d\Omega_{(2,k)}^{2}\, .
\end{equation}



\begin{thebibliography}{99}

\bibitem{Ortin:2015hya}
T.~Ort\'{\i}n,
``Gravity and Strings'', 2nd edition, 
Cambridge University Press, 2015.

\bibitem{Ferrara:1995ih}
S.~Ferrara, R.~Kallosh and A.~Strominger,
``$\mathcal{N}=2$ extremal black holes,''
Phys.\ Rev.\ D {\bf 52} (1995) R5412.
\doi{10.1103/PhysRevD.52.R5412}.
\hepth{9508072}.

\bibitem{Strominger:1996kf}
A.~Strominger,
``Macroscopic Entropy of $\mathcal{N}=2$ Extremal Black Holes,''
Phys.\ Lett.\ B {\bf 383} (1996) 39.
\doi{10.1016/0370-2693(96)00711-3}.
\hepth{9602111}.

\bibitem{Ferrara:1996dd}
S.~Ferrara and R.~Kallosh,
``Supersymmetry and Attractors,''
Phys.\ Rev.\ D {\bf 54} (1996) 1514.
\doi{10.1103/PhysRevD.54.1514}.
\hepth{9602136}.

\bibitem{Ferrara:1996um}
S.~Ferrara and R.~Kallosh,
``Universality of Supersymmetric Attractors,''
Phys.\ Rev.\ D {\bf 54} (1996) 1525.
\doi{10.1103/PhysRevD.54.1525}.
\hepth{9603090}.

\bibitem{Ferrara:1997tw}
S.~Ferrara, G.~W.~Gibbons and R.~Kallosh,
``Black holes and critical points in moduli space,''
Nucl.\ Phys.\ B {\bf 500} (1997) 75.
\doi{10.1016/S0550-3213(97)00324-6}.
[\hepth{9702103}].

\bibitem{Gibbons:1982fy}
G.W.~Gibbons, C.M.~Hull,
``A Bogomolny Bound for General Relativity and Solitons in N=2 Supergravity,''
Phys.\ Lett.\  {\bf 109B} (1982) 190.
\doi{10.1016/0370-2693(82)90751-1}

\bibitem{Tod:1983pm}
K.P.~Tod,
``All Metrics Admitting Supercovariantly Constant Spinors,''
Phys.\ Lett.\  {\bf 121B} (1983) 241.
\doi{10.1016/0370-2693(83)90797-9}.

\bibitem{KowalskiGlikman:1985wi}
J.~Kowalski-Glikman,
``Positive Energy Theorem For Eleven-dimensional Kaluza-klein Supergravity,''
Phys.\ Lett.\  {\bf 166B} (1986) 149.
\doi{10.1016/0370-2693(86)91366-3}.

\bibitem{Perjes:1971gv}
Z.~Perj\'es,
``Solutions of the coupled Einstein Maxwell equations representing the fields of spinning sources,''
Phys.\ Rev.\ Lett.\  {\bf 27} (1971) 1668.
\doi{10.1103/PhysRevLett.27.1668}.

\bibitem{Israel:1972vx}
W.~Israel and G.~A.~Wilson,
``A class of stationary electromagnetic vacuum fields,''
J.\ Math.\ Phys.\  {\bf 13} (1972) 865.
\doi{10.1063/1.1666066}.

\bibitem{Hartle:1972ya}
J.~B.~Hartle and S.~W.~Hawking,
``Solutions of the Einstein-Maxwell equations with many black holes,''
Commun.\ Math.\ Phys.\  {\bf 26} (1972) 87.
\doi{10.1007/BF01645696}.

\bibitem{Majumdar:1947eu}
S.~D.~Majumdar,
``A class of exact solutions of Einstein's field equations,''
Phys.\ Rev.\  {\bf 72} (1947) 390.
\doi{10.1103/PhysRev.72.390}.

\bibitem{kn:P}
 A.~Papapetrou, {\it Proc.~Roy.~Irish.~Acad.}~{\bf A51}  (1947) 191.

\bibitem{Gauntlett:2002nw}
J.~P.~Gauntlett, J.~B.~Gutowski, C.~M.~Hull, S.~Pakis and H.~S.~Reall,
``All supersymmetric solutions of minimal supergravity in five dimensions,''
Class.\ Quant.\ Grav.\  {\bf 20} (2003) 4587.
\doi{10.1088/0264-9381/20/21/005}.
[\hepth{0209114}].

\bibitem{Caldarelli:2003pb}
M.~M.~Caldarelli, D.~Klemm,
``All supersymmetric solutions of N=2, D = 4 gauged supergravity,''
JHEP {\bf 0309} (2003) 019.
\doi{10.1088/1126-6708/2003/09/019}.
[\hepth{0307022}].

\bibitem{Meessen:2006tu}
P.~Meessen, T.~Ort\'{\i}n,
``The supersymmetric configurations of N = 2, d = 4 supergravity coupled to
vector supermultiplets,'' 
Nucl.\ Phys.\ B {\bf 749} (2006) 291.
\doi{10.1016/j.nuclphysb.2006.05.025}.
[\hepth{0603099}].

\bibitem{Huebscher:2006mr}
M.~H\"ubscher, P.~Meessen, T.~Ort\'{\i}n,
``Supersymmetric solutions of N = 2 d = 4 SUGRA: The whole ungauged
shebang,''
Nucl.\ Phys.\  {\bf B759} (2006) 228.
\doi{10.1016/j.nuclphysb.2006.10.004}.
[\hepth{0606281}].

\bibitem{Cacciatori:2008ek}
S.~L.~Cacciatori, D.~Klemm, D.~S.~Mansi, E.~Zorzan,
``All timelike supersymmetric solutions of N=2, D=4 gauged supergravity coupled to abelian vector multiplets,''
JHEP {\bf 0805} (2008) 097.
\doi{10.1088/1126-6708/2008/05/097}.
[\arxiv{0804.0009}]

\bibitem{Klemm:2009uw}
D.~Klemm and E.~Zorzan,
``All null supersymmetric backgrounds of N=2, D=4 gauged supergravity coupled to abelian vector multiplets,''
Class.\ Quant.\ Grav.\  {\bf 26} (2009) 145018.
\doi{10.1088/0264-9381/26/14/145018}.
[\arxiv{0902.4186} [hep-th]].

\bibitem{Klemm:2010mc}
D.~Klemm and E.~Zorzan,
``The timelike half-supersymmetric backgrounds of N=2, D=4 supergravity with Fayet-Iliopoulos gauging,''
Phys.\ Rev.\ D {\bf 82} (2010) 045012.
\doi{10.1103/PhysRevD.82.045012}.
[\arxiv{1003.2974} [hep-th]].

\bibitem{Hubscher:2008yz}
M.~H\"ubscher, P.~Meessen, T.~Ort\'{\i}n, S.~Vaul\`a,
``N=2 Einstein-Yang-Mills's BPS solutions,''
JHEP {\bf 0809 } (2008)  099.
\doi{10.1088/1126-6708/2008/09/099}.
 [\arxiv{0806.1477}].

\bibitem{Meessen:2012sr}
P.~Meessen and T.~Ortin,
``Supersymmetric solutions to gauged N=2 d=4 sugra: the full timelike shebang,''
Nucl.\ Phys.\ B {\bf 863} (2012) 65
\doi{10.1016/j.nuclphysb.2012.05.023}.
[\arxiv{1204.0493}].

\bibitem{Gauntlett:2003fk}
J.~P.~Gauntlett and J.~B.~Gutowski,
``All supersymmetric solutions of minimal gauged supergravity in five-dimensions,''
Phys.\ Rev.\ D {\bf 68} (2003) 105009.
Erratum: [Phys.\ Rev.\ D {\bf 70} (2004) 089901].
\doi{10.1103/PhysRevD.70.089901}, \doi{10.1103/PhysRevD.68.105009}.
[\hepth{0304064}].

\bibitem{Gutowski:2004yv}
J.~B.~Gutowski and H.~S.~Reall,
``General supersymmetric AdS(5) black holes,''
JHEP {\bf 0404} (2004) 048
\doi{10.1088/1126-6708/2004/04/048}
[\hepth{0401129}].

\bibitem{Gauntlett:2004qy}
J.P.~Gauntlett and J.B.~Gutowski,
Phys.\ Rev.\ D {\bf 71} (2005) 045002.
\doi{10.1103/PhysRevD.71.045002}.
[\hepth{0408122}].

\bibitem{Gutowski:2005id}
J.~B.~Gutowski and W.~Sabra,
``General supersymmetric solutions of five-dimensional supergravity,''
JHEP {\bf 0510} (2005) 039
\doi{10.1088/1126-6708/2005/10/039}
[\hepth{0505185}].

\bibitem{Bellorin:2006yr}
J.~Bellor\'{\i}n, P.~Meessen and T.~Ort\'{\i}n,
``All the supersymmetric solutions of N=1,d=5 ungauged supergravity,''
JHEP {\bf 0701} (2007) 020.
\doi{10.1088/1126-6708/2007/01/020}.
[\hepth{0610196}].

\bibitem{Bellorin:2007yp}
J.~Bellor\'{\i}n and T.~Ort\'{\i}n,
``Characterization of all the supersymmetric solutions of gauged N=1, d=5 supergravity,''
JHEP {\bf 0708} (2007) 096
\doi{10.1088/1126-6708/2007/08/096}.
[\arxiv{0705.2567} [hep-th]].

\bibitem{Bellorin:2008we}
J.~Bellor\'{\i}n,
``Supersymmetric solutions of gauged five-dimensional supergravity with general matter couplings,''
Class.\ Quant.\ Grav.\  {\bf 26} (2009) 195012.
\doi{10.1088/0264-9381/26/19/195012}.
[\arxiv{0810.0527} [hep-th]].

\bibitem{Gibbons:1979zt}
G.~W.~Gibbons and S.~W.~Hawking,
``Gravitational Multi - Instantons,''
Phys.\ Lett.\ B {\bf 78} (1978) 430.
\doi{10.1016/0370-2693(78)90478-1}.

\bibitem{Gibbons:1979xm}
G.~W.~Gibbons and S.~W.~Hawking,
``Classification of Gravitational Instanton Symmetries,''
Commun.\ Math.\ Phys.\  {\bf 66} (1979) 291.
\doi{10.1007/BF01197189}.

\bibitem{Meessen:2015enl}
P.~Meessen, T.~Ort\'{\i}n and P.~Fern\'andez-Ram\'{\i}rez,
``Non-Abelian, supersymmetric black holes and strings in 5 dimensions,''
JHEP {\bf 1603} (2016) 112.
\doi{10.1007/JHEP03(2016)112}.
[\arxiv{1512.07131} [hep-th]].

\bibitem{kn:KronheimerMScThesis}
P.B.~Kronheimer, 
``Monopoles and Taub-NUT spaces,'' 
M.Sc.~Dissertation, Oxford University, 1995.

\bibitem{Bogomolny:1975de}
E.~B.~Bogomol'nyi,
``Stability of Classical Solutions,''
Sov.\ J.\ Nucl.\ Phys.\  {\bf 24} (1976) 449
[Yad.\ Fiz.\  {\bf 24} (1976) 861].

\bibitem{Protogenov:1977tq}
A.~P.~Protogenov,
``Exact Classical Solutions of Yang-Mills Sourceless Equations,''
Phys.\ Lett.\ B {\bf 67} (1977) 62.
\doi{10.1016/0370-2693(77)90806-1}.

\bibitem{Belavin:1975fg}
A.A.~Belavin, A.M.~Polyakov, A.S.~Schwartz and Y.S.~Tyupkin,
``Pseudoparticle solutions of the Yang-Mills equations,''
Phys.\ Lett.\ B {\bf 59} (1975) 85.
\doi{10.1016/0370-2693(75)90163-X}.

\bibitem{Bueno:2015wva}
P.~Bueno, P.~Meessen, T.~Ort\'{\i}n and P.~F.~Ram\'{\i}rez,
``Resolution of SU(2) monopole singularities by oxidation,''
Phys.\ Lett.\ B {\bf 746} (2015) 109.
\doi{10.1016/j.physletb.2015.04.065}.
[\arxiv{1503.01044} [hep-th]].

\bibitem{Meessen:2008kb}
P.~Meessen,
``Supersymmetric coloured/hairy black holes,''
Phys.\ Lett.\ B {\bf 665} (2008) 388.
\doi{10.1016/j.physletb.2008.06.035}.
[\arxiv{0803.0684} [hep-th]].

\bibitem{Meessen:2015nla}
P.~Meessen and T.~Ort\'{\i}n,
``$ \mathcal{N}=2 $ super-EYM coloured black holes from defective Lax matrices,''
JHEP {\bf 1504} (2015) 100.
\doi{10.1007/JHEP04(2015)100}.
[\arxiv{1501.02078} [hep-th]].

\bibitem{Figueras:2006xx}
P.~Figueras, C.~A.~R.~Herdeiro and F.~Paccetti Correia,
``On a class of 4D Kahler bases and AdS(5) supersymmetric Black Holes,''
JHEP {\bf 0611} (2006) 036
\doi{10.1088/1126-6708/2006/11/036}
[\hepth{0608201}].

\bibitem{Cassani:2015upa}
D.~Cassani, J.~Lorenzen and D.~Martelli,
``Comments on supersymmetric solutions of minimal gauged supergravity in five dimensions,''
Class.\ Quant.\ Grav.\  {\bf 33} (2016) no.11,  115013
\doi{10.1088/0264-9381/33/11/115013}
[\arxiv{1510.01380} [hep-th]].

\bibitem{London:1995ib}
L.~A.~J.~London,
``Arbitrary dimensional cosmological multi - black holes,''
Nucl.\ Phys.\ B {\bf 434} (1995) 709.
\doi{10.1016/0550-3213(94)00511-C}.

\bibitem{Gutowski:2004ez}
J.~B.~Gutowski and H.~S.~Reall,
``Supersymmetric AdS(5) black holes,''
JHEP {\bf 0402} (2004) 006
\doi{10.1088/1126-6708/2004/02/006}
[\hepth{0401042}].

\bibitem{Chimento:2016run}
S.~Chimento and T.~Ort\'{\i}n,
``On 2-dimensional K\"ahler metrics with one holomorphic isometry,''
\arxiv{1610.02078} [hep-th].

\bibitem{Bergshoeff:2004kh}
E.~Bergshoeff, S.~Cucu, T.~de Wit, J.~Gheerardyn, S.~Vandoren and A.~Van Proeyen,
``N = 2 supergravity in five-dimensions revisited,''
Class.\ Quant.\ Grav.\  {\bf 21} (2004) 3015.
[Class.\ Quant.\ Grav.\  {\bf 23} (2006) 7149].
\doi{10.1088/0264-9381/23/23/C01}, \doi{10.1088/0264-9381/21/12/013}.
[\hepth{0403045}].

\bibitem{kn:Joyce}
D.~Joyce,
``Compact Manifolds with Special Holonomy,''
Oxford University Press, U.K. (2000)

\bibitem{Gibbons:2011sg}
G.~W.~Gibbons,
``Anti-de-Sitter spacetime and its uses,''
\arxiv{1110.1206} [hep-th].

\bibitem{Trautman:1977im}
A.~Trautman,
``Solutions of the Maxwell and Yang-Mills Equations Associated with Hopf Fibrings,''
Int.\ J.\ Theor.\ Phys.\  {\bf 16} (1977) 561.
\doi{10.1007/BF01811088}

\bibitem{Penrose:1965am}
R.~Penrose,
``Zero rest mass fields including gravitation: Asymptotic behavior,''
Proc.\ Roy.\ Soc.\ Lond.\ A {\bf 284} (1965) 159.
\doi{10.1098/rspa.1965.0058}.

\bibitem{Ashtekar:1999jx}
A.~Ashtekar and S.~Das,
``Asymptotically Anti-de Sitter space-times: Conserved quantities,''
Class.\ Quant.\ Grav.\  {\bf 17} (2000) L17.
\doi{10.1088/0264-9381/17/2/101}.
[\hepth{9911230}].

\bibitem{Klemm:2000nj}
D.~Klemm and W.~A.~Sabra,
``Supersymmetry of black strings in D = 5 gauged supergravities,''
Phys.\ Rev.\ D {\bf 62} (2000) 024003.
\doi{10.1103/PhysRevD.62.024003}.
[\hepth{0001131}].

\bibitem{Godel:1949ga}
K.~G\"odel,
``An Example of a new type of cosmological solutions of Einstein's field equations of graviation,''
Rev.\ Mod.\ Phys.\  {\bf 21} (1949) 447.
\doi{10.1103/RevModPhys.21.447}.

\bibitem{Tseytlin:1996as}
A.~A.~Tseytlin,
``Extreme dyonic black holes in string theory,''
Mod.\ Phys.\ Lett.\ A {\bf 11} (1996) 689.
\doi{10.1142/S0217732396000709}.
[\hepth{9601177}].

\bibitem{Meessen:2004mh}
P.~Meessen and T.~Ort\'{\i}n,
``G\"odel space-times, Abelian instantons, the graviphoton background and other flacuum solutions,''
Nucl.\ Phys.\ B {\bf 684} (2004) 235.
\doi{10.1016/j.nuclphysb.2004.02.020}.
[\hepth{0401005}].

\bibitem{Gaddam:2014mna}
N.~Gaddam, A.~Gnecchi, S.~Vandoren and O.~Varela,
``Rholography, Black Holes and Scherk-Schwarz,''
JHEP {\bf 1506} (2015) 058.
\doi{10.1007/JHEP06(2015)058}.
[\arxiv{1412.7325} [hep-th]].

\bibitem{Chong:2005da}
Z.~W.~Chong, M.~Cvetic, H.~Lu and C.~N.~Pope,
``Five-dimensional gauged supergravity black holes with independent rotation parameters,''
Phys.\ Rev.\ D {\bf 72} (2005) 041901.
\doi{10.1103/PhysRevD.72.041901}.
[\hepth{0505112}].

\bibitem{Kunduri:2006ek}
H.~K.~Kunduri, J.~Lucietti and H.~S.~Reall,
``Supersymmetric multi-charge AdS(5) black holes,''
JHEP {\bf 0604} (2006) 036.
\doi{10.1088/1126-6708/2006/04/036}.
[\hepth{0601156}].

\bibitem{kn:CO}
S.~Chimento and T.~Ort\'{\i}n,
work in progress.

\bibitem{Dunajski:2013qc}
M.~Dunajski, J.~Gutowski and W.~Sabra,
``Enhanced Euclidean supersymmetry, 11D supergravity and $SU(\infty)$ Toda equation,''
JHEP {\bf 1310} (2013) 089.
\doi{10.1007/JHEP10(2013)089}.
[\arxiv{1301.1896} [hep-th]].

\bibitem{Ketov:2001gq}
S.~V.~Ketov,
``Universal hypermultiplet metrics,''
Nucl.\ Phys.\ B {\bf 604} (2001) 256.
\doi{10.1016/S0550-3213(01)00184-5}.
[\hepth{0102099}].

\bibitem{Ashtekar:1984zz}
A.~Ashtekar and A.~Magnon,
``Asymptotically anti-de Sitter space-times,''
Class.\ Quant.\ Grav.\  {\bf 1} (1984) L39.
\doi{10.1088/0264-9381/1/4/002}.





\end{thebibliography}
\end{document}